\documentclass[aps,prd,10pt,notitlepage,nofootinbib,superscriptaddress,showkeys,showpacs]{revtex4-1}
\pdfoutput=1

\usepackage{amstext,amsmath,amssymb,amsfonts,bbm}
\usepackage[latin1]{inputenc}
\usepackage{epsfig}
\usepackage{hyperref}
\usepackage{amsthm}
\usepackage{subfigure}
\usepackage{color}
\usepackage{multirow}

\usepackage{graphicx}

\usepackage{latexsym}

\usepackage{cancel}

\newcommand{\bea}{\begin{eqnarray}}	
\newcommand{\eea}{\end{eqnarray}}
\newcommand{\be}{\begin{equation}}	
\newcommand{\ee}{\end{equation}}
\newcommand{\beq}{\begin{equation}}	
\newcommand{\eeq}{\end{equation}}

\newcommand{\cA}{ \mathcal{A} } 
\newcommand{\cB}{\mathcal{B}}

\newcommand{\cV}{{\mathcal V}}
\newcommand{\cZ}{ \mathcal{Z} }

\newcommand{\Z}{{\mathbb Z}}
\newcommand{\C}{{\mathbb C}}
\newcommand{\R}{{\mathbb R}}

\newcommand{\inter}{{\rm int\,}}

\newcommand{\bee}{{\mathbf{b}}}
\newcommand{\Tr}{{\rm Tr}}

\newcommand{\ft}{ {\textstyle{\frac13}}}
\newcommand{\fd}{ {\textstyle{\frac1d}}}

\newcommand{\bdel}{{\boldsymbol{\delta}}}
\newcommand{\Sym}{ {\rm Sym} }

\newcommand{\wthet}{ \widetilde{\Theta} } 
\newcommand{\inc}{
\includegraphics[angle=0, width=1.2cm, height=0.5cm]{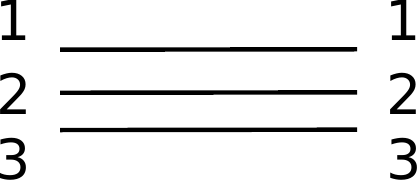}}
\newcommand{\incp}{
\includegraphics[angle=0, width=1.2cm, height=0.5cm]{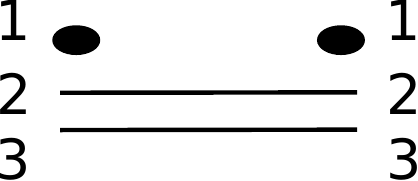}}
\newcommand{\incpp}{
\includegraphics[angle=0, width=1.2cm, height=0.5cm]{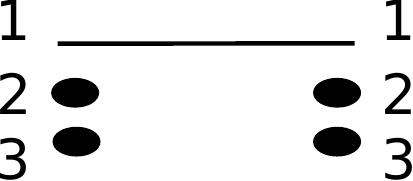}}

\newcommand{\incPp}{
\includegraphics[angle=0, width=1.2cm, height=0.5cm]{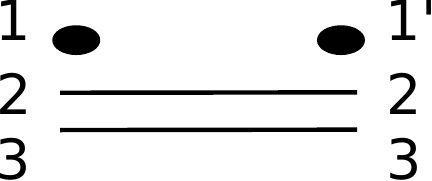}}
\newcommand{\incPpp}{
\includegraphics[angle=0, width=1.2cm, height=0.5cm]{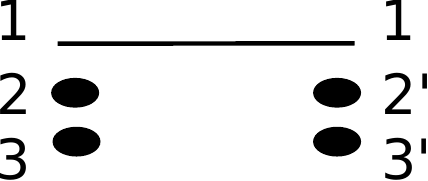}}
\newcommand{\incPppp}{
\includegraphics[angle=0, width=1.2cm, height=0.5cm]{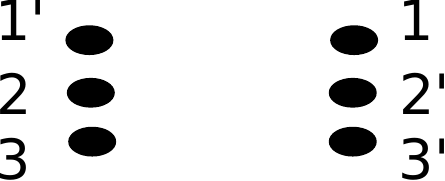}}
\newcommand{\incPHppp}{
\includegraphics[angle=0, width=1.2cm, height=0.5cm]{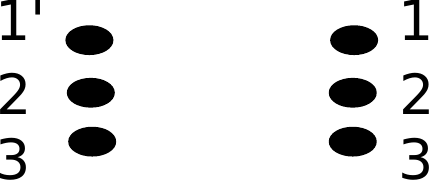}}
\newcommand{\incPHIppp}{
\includegraphics[angle=0, width=1.2cm, height=0.5cm]{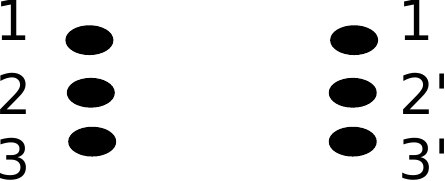}}
\newcommand{\incPtwo}{
\includegraphics[angle=0, width=1.2cm, height=0.5cm]{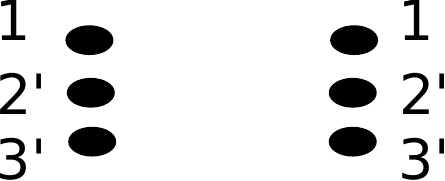}}
\newcommand{\incPthree}{
\includegraphics[angle=0, width=1.2cm, height=0.5cm]{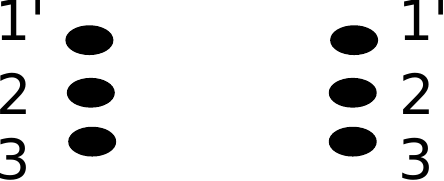}}

\newcommand{\intPp}{
\includegraphics[angle=0, width=1.2cm, height=0.5cm]{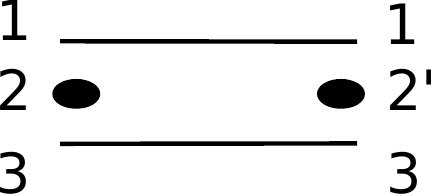}}
\newcommand{\intPpp}{
\includegraphics[angle=0, width=1.2cm, height=0.5cm]{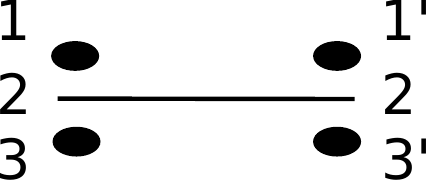}}
\newcommand{\intPppp}{
\includegraphics[angle=0, width=1.2cm, height=0.5cm]{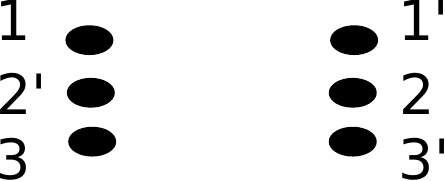}}

\newcommand{\intPHpp}{
\includegraphics[angle=0, width=1.2cm, height=0.5cm]{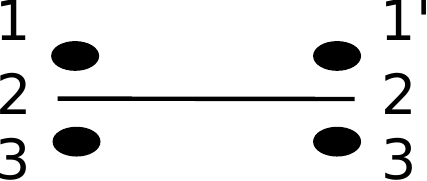}}

\newcommand{\intpq}{
\includegraphics[angle=0, width=1.2cm, height=0.5cm]{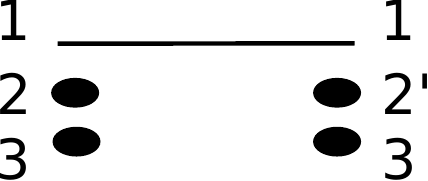}}

\newcommand{\intmp}{
\includegraphics[angle=0, width=1.2cm, height=0.5cm]{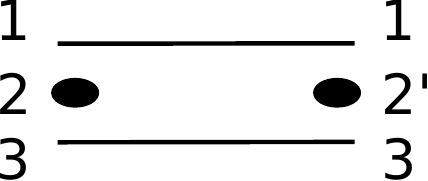}}

\newcommand{\intpr}{
\includegraphics[angle=0, width=1.2cm, height=0.5cm]{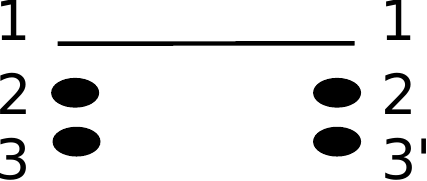}}

\newcommand{\intppr}{
\includegraphics[angle=0, width=1.2cm, height=0.5cm]{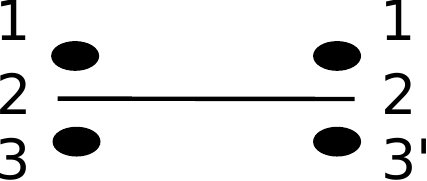}}

\newcommand{\intpqr}{
\includegraphics[angle=0, width=1.2cm, height=0.5cm]{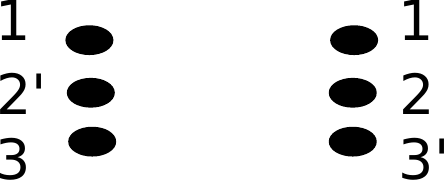}}

\newcommand{\intope}{
\includegraphics[angle=0, width=1.2cm, height=0.5cm]{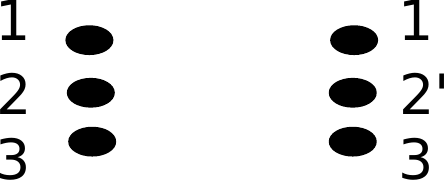}}

\newcommand{\intobe}{
\includegraphics[angle=0, width=1.2cm, height=0.5cm]{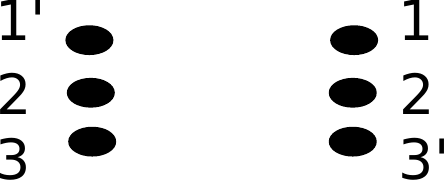}}

\newcommand{\vex}{
\includegraphics[angle=0, width=1.2cm, height=1cm]{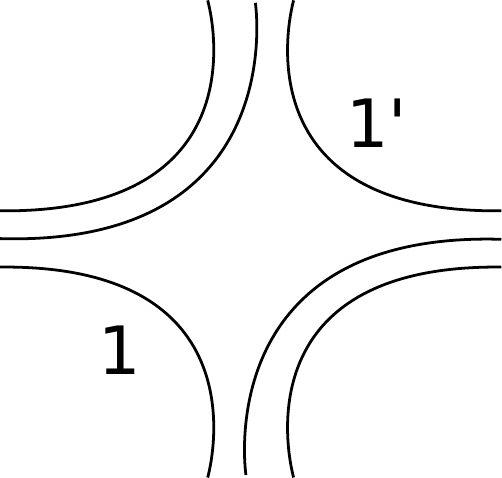}}

\newcommand{\vexpq}{
\includegraphics[angle=0, width=1.2cm, height=1cm]{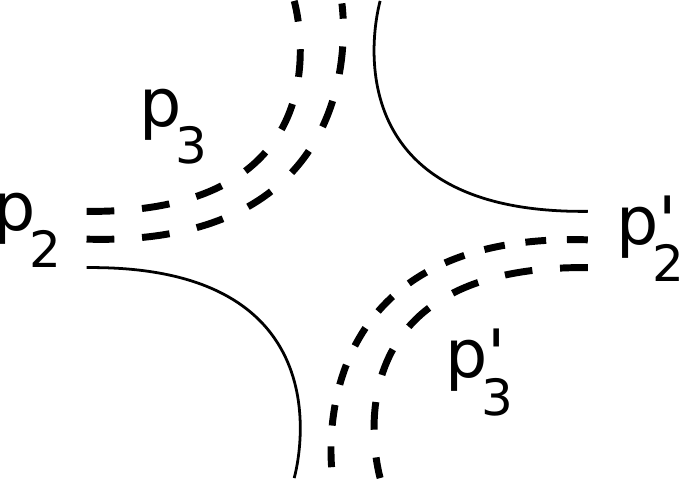}}

\newcommand{\vone}{
\includegraphics[angle=0, width=1.2cm, height=1cm]{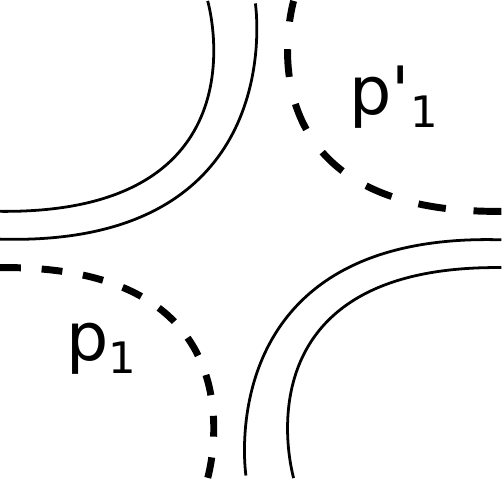}}

\newcommand{\incross}{
\includegraphics[angle=0, width=1.2cm, height=1cm]{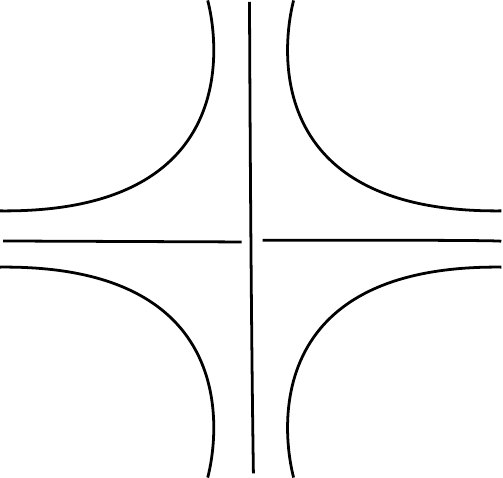}}

\newcommand{\vontr}{
\includegraphics[angle=0, width=1.2cm, height=1cm]{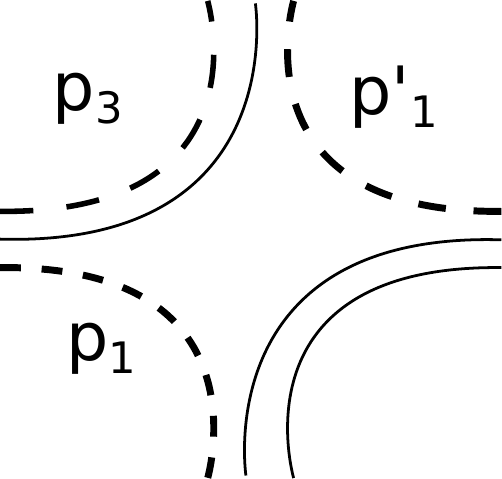}}

\newcommand{\vosec}{
\includegraphics[angle=0, width=1.2cm, height=1.1cm]{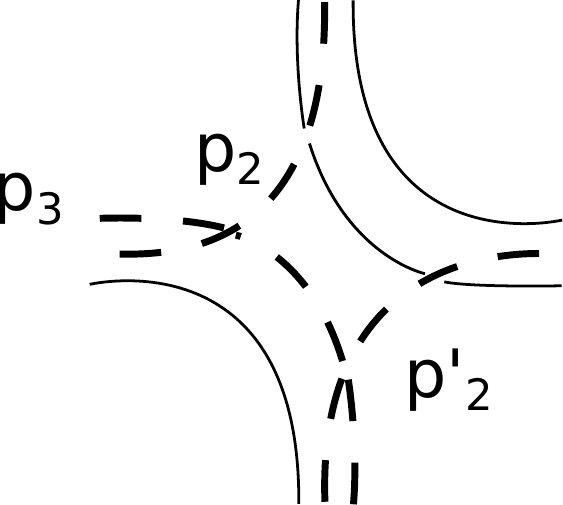}}

\newcommand{\secvex}{
\includegraphics[angle=0, width=1.2cm, height=1.1cm]{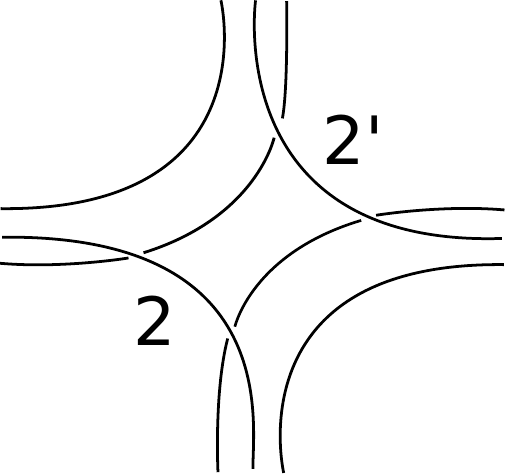}}

\newcommand{\trvex}{
\includegraphics[angle=0, width=1.2cm, height=1.1cm]{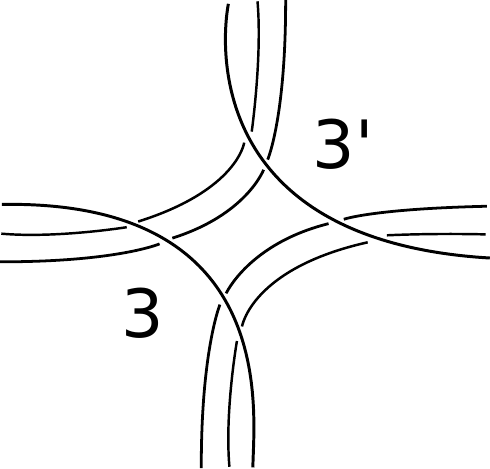}}

\begin{document}

\title{Functional Renormalisation Group Approach
for Tensorial Group Field Theory: \\
 a Rank-3 Model}

\author{Dario Benedetti}\email{dario.benedetti@aei.mpg.de}

\affiliation{Max Planck Institute for Gravitational Physics, Albert Einstein Institute\\
 Am M\"uhlenberg 1, 14476, Potsdam, Germany }

\author{Joseph Ben Geloun}\email{jbengeloun@aei.mpg.de}

\affiliation{Max Planck Institute for Gravitational Physics, Albert Einstein Institute\\
 Am M\"uhlenberg 1, 14476, Potsdam, Germany }

\affiliation{International Chair in Mathematical Physics and Applications, 
ICMPA-UNESCO Chair, 072BP50, Cotonou, Rep. of Benin }

\author{Daniele Oriti}\email{daniele.oriti@aei.mpg.de}

\affiliation{Max Planck Institute for Gravitational Physics, Albert Einstein Institute\\
 Am M\"uhlenberg 1, 14476, Potsdam, Germany }

\begin{abstract}

We set up the Functional Renormalisation Group formalism for Tensorial
Group Field Theory in full generality. We then apply it to a rank-3 model over $U(1)^3$, endowed with a kinetic term linear in the momenta and with nonlocal 
interactions. The system of FRG equations turns out to be non-autonomous in the RG flow parameter. This feature is explained by the existence of a hidden scale, the radius of the group manifold. 
We investigate in detail the opposite regimes of large cut-off (UV) and small cut-off (IR) of the FRG equations, where the system becomes autonomous, and we
find, in both case, Gaussian and non-Gaussian fixed points. We derive and interpret the critical exponents and flow diagrams associated with these fixed points, and discuss how the UV and IR regimes are matched. 
Finally, we discuss the evidence for a phase transition from a symmetric phase to a broken or condensed phase, from an RG perspective, finding that this seems to exist only in the approximate regime of very large radius of the group manifold, as to be expected for systems on compact manifolds.

\

\end{abstract}

\maketitle

\tableofcontents

\section{Introduction}

Group field theories \cite{GFTreviews} (GFTs) are a new class of quantum field theories, defined on group manifolds and characterised by a specific form of non-locality in their interactions, giving their Feynman diagrams the combinatorial structure of cellular complexes, rather than graphs. It has found traction, beyond its intrinsic mathematical interest, in recent developments in quantum gravity research. Technically, they are quantum field theories on group manifolds characterised by interaction terms involving combinatorially non-trivial patterns of convolutions of field arguments, in turn giving the Feynman diagrams of the theory the structure of cellular complexes rather than simple graphs. 

Historically, this type of models arose from the attempts to generalise matrix models of 2d quantum gravity \cite{Di Francesco:1993nw} to higher dimensions, in the form of tensor models \cite{tensor}. Soon afterwards \cite{boulatov}, the structure of tensor models was enriched with group-theoretic data in order to obtain models whose Feynman amplitude matched state sum models of topological field theories. This gave the first examples of group field theories as currently understood. It also established a solid link \cite{carloPR} with loop quantum gravity \cite{LQG} (LQG), in that the additional data implied a close relation between the quantum states of group field theories and those found in the canonical approach. This link became the focus of later research, when it was understood \cite{carlomike} that group field theories provide a complete definition of spin foam models \cite{SF}, being developed as a covariant version of loop quantum gravity. Finally, the relation between group field theories and discrete gravity, already evident in their origin in tensor models, becomes even more strict thanks to results showing the appearance of the Regge action in semiclassical analyses of spin foam amplitudes (for the many results in this direction, see for example, \cite{SFasympt} and references therein), and, more recently, the general possibility to recast group field theory amplitudes as (non-commutative) simplicial gravity path integrals \cite{danielearistide}, which also clarify the appearance of the Regge action from a different angle. Group field theories sit then at the crossroad of several approaches to quantum gravity, as a 2nd quantised framework for loop quantum gravity degrees of freedom \cite{GFT-LQG} as well as an enrichment of tensor models. 

Tensor models represent not only the historical origin of group field theories, but also their mathematical backbone. In fact, they have witnessed a powerful resurgence in recent years \cite{Rivasseau:2011hm}, which has directly benefitted GFTs as well. New tools for analysing the combinatorics of GFT/tensor Feynman diagrams have been developed 
\cite{coloured}, leading to several key results on their large-$N$ expansion \cite{largeN}, 
double scaling limits \cite{doublescal}, 
critical behaviour \cite{criticalTensor} and universality \cite{universalityTensor}. 
For a recent review, see \cite{tensorNew}. Several of these results have been extended to full-fledged group field theories as well \cite{GFTscaling, GFTphase}. 
In particular, developments in tensor models suggest a definition of the GFT theory space \cite{vincentTheorySpace} that is specific yet general enough to define an interesting subclass of GFTs, called Tensorial GFTs (TGFTs) in the literature, on which we focus in this article as well. TGFTs can also be seen, in fact, as tensor models enriched with group-theoretic data, turning hem into proper quantum field theories. We are going to rely on this perspective in this paper.

One main advantage of group field theories, from the perspective of related quantum gravity approaches, is that they allow the use of standard quantum field theory tools in a background independent context. These QFT tools become in particular important for tackling the issue of the emergence of continuum spacetime and gravity from quantum gravity dynamics. 

This problem is of course a crucial one in background independent quantum gravity approaches whose basic degrees of freedom are, to a varying extent, pre-geometric and not spatio-temporal \cite{danieleemergence}. It requires controlling the quantum dynamics of many such interacting degrees of freedom and extracting their effective collective physics. In particular, it implies unraveling the macroscopic phase structure of the given quantum gravity model and the phase transitions it may undergo. One of the continuum phases should of course correspond to a geometric regime of the theory, in which standard General Relativity applies, at least in some approximation. The putative phase transition separating such geometric phase from a generic non-geometric one has been called geometrogenesis \cite{danieleemergence}, and its cosmological interpretation \cite{LeeJoaoContaldi} (as what replaces the classical big bang singularity in a full quantum gravity framework) has received some indirect support from recent work on GFT condensates \cite{GFTcondensate}. Indeed, one possible scenario identifies the geometrogenesis transition in a condensation of GFT quanta, with the geometric phase characterised, in mean field approximation, by a non-zero expectation value of the GFT field operator, similarly to what happens in standard Bose condensates. This idea has been also suggested from a canonical LQG perspective \cite{Tim-DGP,HannoTim}. The same issue has of course been tackled in related approaches, most notably causal dynamical triangulations \cite{CDTphasetrans}, with a different perspective, even if ending up sometimes with a similar picture \cite{dariojoe,jacub}. 

In addressing the problem of the continuum, and of extracting the phase diagram of the quantum gravity theory, renormalisation group methods come of course to prominence. They are being developed as well in the context of spin foam models from a lattice gauge theory perspective \cite{biancabenny}, but GFTs allow, as mentioned, a most direct application of renormalisation formalisms to quantum gravity, thanks to their more or less standard QFT nature. In the GFT context, the renormalisation group has first of all the role of testing the consistency of the perturbative expansion that usually defines the quantum theory, and indeed recent years have witnessed a flourishing of studies of perturbative renormalisability of GFT models \cite{GFTrenorm}\cite{Carrozza:2013mna}, mainly in the context of Tensorial GFTs. Next, it is employed indeed to study the flow of GFT coupling constants and to map out the phase diagram of the same models. Beta function computations have already been performed for some GFT models \cite{BenGeloun:2012pu,josephBeta, Samary:2013xla, Sylvain} using perturbative methods, and suggest that asymptotic freedom in GFT could be generic, even though it remains difficult to establish beyond doubt.

Functional renormalisation group (FRG) methods \cite{Berges:2000ew,Bagnuls:2000ae,Delamotte-review} are another powerful tool for the study of the phase structure of quantum field theories, in particular in situations where perturbation theory is expected not to apply, for example in the vicinity of non-trivial fixed points. 
Beside being the key technology for the asymptotic safety approach to quantum gravity \cite{AS}, the FRG may provide an important asset to group field theories. The key difference of GFTs with respect to usual field theories is represented by their combinatorially non-trivial interactions. This feature is shared by the much simpler matrix models for 2d gravity, the renormalisation analysis of which was initiated in \cite{Brezin:1992yc}, adapting the Wilsonian idea of integrating out a momentum shell to the matrix case.  A first important application of FRG methods to matrix models was recently done in \cite{AstridTim} and further in \cite{AstridTim2}, which represented a nice display of the power of such methods to reproduce, at least qualitatively, known results about phase transitions in such models. In the same papers, it was suggested that similar methods could be applied to tensor models.

In this paper, we pick up on that suggestion, develop for the first time the general formalism of the functional renormalisation group in the context of (tensorial) group field theories, and apply it to one specific example, to illustrate the general procedure. A difference in perspective, with respect to \cite{AstridTim}, should be noted: while taking heavily into account, as appropriate, the tensorial aspects of TGFTs, notably the peculiar combinatorial structure of interactions, we rely mainly on their quantum field theory aspects to set up and apply our FRG formalism.

More precisely, we do the following. We focus on Tensorial GFTs, treating them as standard QFTs on a compact manifolds, but taking into account, as appropriate and necessary, the tensorial aspect of their interactions. To this class of field theories, we adapt and set up the FRG formalism. That is, we write down the GFT effective action, having implemented a suitable cut-off function, and obtain the GFT version of the standard Wetterich equation \cite{Wetterich:1992yh,Morris:1993qb}. We clarify the canonical dimensions of GFT fields and couplings and the role of truncations that are necessary to solve the Wetterich equation. We also explain why one should expect a non-autonomous system of beta functions for this type of field theories. This is one main result of the present work. 
We then move to the analysis of a concrete example of Tensorial GFT. We choose to study a 3-dimensional abelian model with non-trivial but linear kinetic term, in its simplest truncation. For this model, which is already computationally non-trivial, we derive and analyse the explicit  equations for the beta functions. We analyse in particular its UV (large-cut-off) regime as well as its IR (small cut-off) regime. In both cases, we identify the fixed points and critical directions of the renormalisation group flow. In the UV regime, we are able to confirm the perturbative result of the asymptotic freedom of the model, as well as finding evidence of a non-Gaussian fixed point. In the IR sector, where we find the same type of fixed points, we discuss the issue of existence of a phase transition separating a symmetric phase from a broken or condensate phase, from an RG perspective. We find that the RG flow equations as such do not give any strong evidence for its existence, without further approximations, for the model considered. However, they suggest that such phase transition should be found in the regime of very large radius of the group manifold. This detailed analysis provides a template of what one should expect from the FRG analysis of GFTs more generally.
We conclude by a brief summary of our results.
Some appendices collect further details on the calculation of the flow equations and on other relevant aspects (e.g. the definition of canonical dimensions for fields and coupling constants) of our analysis.

\section{Tensorial group field theory models}
\label{sect:tens}

GFTs are, as mentioned, QFTs on group manifolds characterised by a peculiar combinatorial pattern of convolutions of field arguments in the interactions. Tensorial GFT models (TGFTs) are those GFTs in which a tensorial transformation property is required for the fields, with specific invariances in the interaction terms. Specific models are defined by a precise choice of group manifold, kinetic and interaction kernels, as well as by the detail of convolution patterns. In this section, we give first a general definition of GFT models, and for the TGFT class. Then, we introduce the particular example we will analyse from an FRG point of view in the following: an abelian TGFT with linear kinetic term.     

\subsection{Group field theory formalism}
\label{subsect:tens}

Consider a field $\phi$ over $d$ copies of a 
compact Lie group $G$, $\varphi: G^d \to \mathbb{K}$,
$\mathbb{K}=\C,\R$. Models defined on a non-compact group manifold, notably the Lorentz group, have been studied but will not concern us here. Rather, we will give a few more details on the cases corresponding to $G=U(1)$ or $G=SU(2)$, while trying at the same time to keep the presentation general as, indeed, our FRG set up does not depend so much on specific choices. We do not assume any special symmetry for the fields $\phi$. Using the group-theoretic generalisation of standard Fourier decomposition,\footnote{Another non-commutative generalisation of the standard Fourier transform \cite{QuantFlux}, leads to a formulation of group field theories (and spin foam models) in terms of Lie algebra variables \cite{danielearistide}, indeed classically conjugate to the group variables. We will not use it in the following, though.} namely Peter-Weyl decomposition for compact Lie groups (or the Plancherel decomposition in the non-compact case),
we write
\beq
\phi(g_1,\dots, g_d) = \sum_{{\bf P}} \phi_{{\bf P}}  \prod_{i=1}^d D^{p_i}(g_i) \qquad .
\eeq
In this formula, ${\bf P} = (p_1, \dots, p_d)$ and the functions $D^{p_i}(g_i)$ form a complete and orthonormal basis of functions on the group, with the $p_i$ being a set of labels characterising them. In particular, the functions  $D^{p_i}(g_i)$ are the representation matrices, in a basis of complete eigenstates of Lie algebra operators with set of eigenvalues $p_i$, of the group elements $g_i$. 
In the case of $G=SU(2)$, the $D$-functions are the well-known Wigner functions $D_{mn}^J(g) = \langle J m | g^{(J)} | J n \rangle$, labelled by the eigenvalues $J(J+1)$ of the quadratic Casimir, with $2 J \in \mathbb{N}$, and by the eigenvalues $m,n \in ( - J, +J) $ of a given generator of the Lie algebra. In the simpler case of  $G=U(1)$, for group elements $g = e^{i \theta}$, $\theta = [0, 2\pi)$, the representation functions take the form $D^n(g)= e^{i n \theta}$, with $n\in\Z$. The representation functions play here the role of plane waves, with the representation labels identifying the analogues of momentum modes for the fields.

A generic action $S[\phi]$ for a GFT model written in the ${\bf P}$'s
basis is  determined by a choice of  kinetic  and interaction kernels:
\bea
\label{eq:actiond}
&&
S[\phi]= Z \, \Tr_2 (\bar\phi \cdot K \cdot \phi) 
+ m \Tr_2 (\bar \phi \phi) + S^{\inter}[\phi]\,, 
\cr\cr
&&
\Tr_2 (\bar\phi \cdot K \cdot \phi) =
\sum_{{\bf P}, \, \tilde{{\bf P}}} \bar\phi_{{\bf P}} \, K({\bf P};\tilde{{\bf P}})\, \phi_{\tilde{{\bf P}} } \,, 
\qquad 
\Tr_{2}(\phi^2) = \sum_{{\bf P}} \bar\phi_{{\bf P}}\phi_{{\bf P}}\,, 
 \cr\cr
&& S^{\inter}[\phi]=  \sum_{n_b} 
\lambda_{n_b}  \Tr_{n_b}(\cV_{n_b}\cdot \phi^{n_b})\,,
\eea
where $\bar\phi$ denotes the complex conjugate of $\phi$. The notation
$\Tr_n$ refers to a summation over the values ${\bf P}$ for $n$ GFT fields.
$K$, respectively $\cV_n$, are kernels convoluting the  arguments of $2$, respectively $n$, fields $\phi$.
The symbol $b$ here parametrizes the type of contraction (or convolution) one is performing
on the $n$ fields, i.e. their specific combinatorial pattern of tracing over the set of indices ${\bf P}$. In the specific instance $n=2$, there is a single possibility and one has $\Tr_2(\phi^2)$. 
$Z$ and $m$ are wave function and
mass couplings, $\lambda_{n_b}$ are interaction coupling constants. 
The sum over $n_b$ is over a finite number of interactions.

Given this action, the partition function is defined as usual:

\be
\label{eq:partLam}
\cZ[J] =
e^{W[J]}
 = \int_M d\phi\,  e^{- S[\phi] + \Tr_2 (J \cdot \phi)} \,,
\ee
where $J$ is a rank $d$ source term, $\Tr_2 (J \cdot \phi) = \sum_{{\bf P}} J_{{\bf P}}\phi_{{\bf P}}$, and $M$ is a cut-off
on the modes $\phi_{\bf P}$, i.e. $|p_i|\leq M$.

The tensorial subclass of GFTs corresponds to the case in which the field $\phi$ transforms as a proper tensor under some unitary (or orthogonal, in the real case) group of symmetries $\mathfrak{U}^{\times d}$, and the interactions correspond to invariant field monomials under the same symmetry group. In particular, in the presence of a cutoff in modes, and with the further restriction to positive values for the labels, i.e. for $p_i=1,2,...M$, the models in \eqref{eq:actiond} can be seen as a straightforward generalization of matrix models.
In such case, dealing with a tensor $\phi_{{\bf P}}$, its indices are all distinct
and can be each associated with a color, and the pattern of contractions in each interaction can be represented by a colored graph, each corresponding to a $U(M)$ invariant monomial in the complex case \cite{Gurau:2011tj,Gurau:2012ix} or
a $O(M)$ invariant in the real case.

\subsection{A rank 3 $O(2M+1)$ invariant tensorial model}
\label{subsect:onmod}

Within the general class of Tensorial GFTs introduced above, we will consider a specific example, that we introduce here, also to clarify some details of the general formalism.
It is a model for a rank $d=3$ real tensor field $\phi:U(1)^3\to \R$.
After mode decomposition using the simple plane waves $D^{p_s}(g_s)=e^{i p_s \theta_s}$, where $\theta_s \in [0,2\pi)$ parametrizes $S^1 \sim U(1)$, one gets field modes $ \phi_{\bf P}= \phi_{p_1,p_2,p_3} 
=\phi_{123}$, $p_i \in \Z$.

\begin{figure}[h]
 \centering
     \begin{minipage}{.5\textwidth}
\includegraphics[angle=0, width=1.5cm, height=1.5cm]{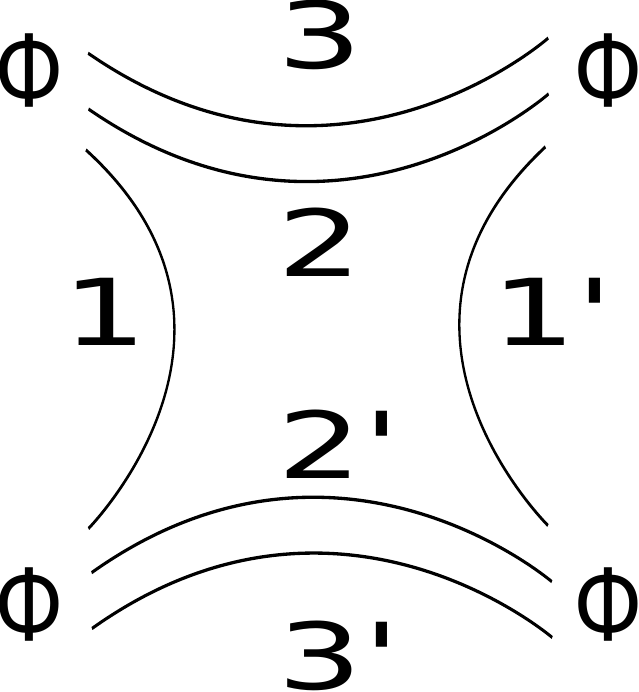} 
\caption{ {\small  The vertex $\Tr_{4;1}(\phi^4)$.  }} 
\label{fig:4vertex}
\end{minipage}
\end{figure}

The model is defined by the action
\be\label{simact}
S[\phi]= \frac{Z}{2} \, \Tr_2 (\phi \cdot K \cdot \phi) 
+ \frac{m}{2} \Tr_2 (  \phi^2) + S^{\inter}[\phi]\,.
\ee
The kernel of the kinetic and interaction terms encoding the dynamics are as follows:
\bea
\label{eq:3daction}
&&
K(\{p_i\};\{p_i'\}) = \bdel_{p_i,p_i'} (\ft \sum_{i=1}^3 |p_i|)  \,,
\qquad 
\bdel_{p_i,p_i'} := \prod_{i=1}^3 \delta_{p_i,p_i'} \,,
\qquad 
\Tr_{2}(\phi^2)= \sum_{p_i \in \Z} \phi_{123} ^2 \,,
\cr\cr
&&
S^{\inter}[\phi]=  \frac{\lambda}{4} \Tr_{4}(\phi^4)
 :=  \frac{\lambda}{4} \Big( \Tr_{4;1} (\phi^4) + \Sym (1 \to 2 \to 3)\Big) 
\,, \qquad 
\Tr_{4;1}(\phi^4) = \sum_{p_i, p_i' \in \Z} 
\phi_{123} \,\phi_{1'23} \,\phi_{1'2'3'} \,\phi_{12'3'} \,.
\eea
The kernel term $Z\Tr_2 (\phi \cdot K\cdot \phi)/2$ plays the role of a kinetic term. 
Considering a Laplacian $\Delta$ on $U(1)$, the kinetic
term is built from a sum of square roots of the eigenvalues $|p_i|=\sqrt{p_i^2}$ of three
Laplacian operators over the three copies of $U(1)$. 
The pattern of the tensor contraction in the interaction term $\Tr_{4;1}(\phi^4)$ can be represented by Figure \ref{fig:4vertex}. Each line element
in the figure represents an index contracted with another
index in another tensor. 
The rest of the terms appearing in $\Sym (1 \to 2 \to 3)$ can be simply inferred by color permutation symmetry from this first term. 

\begin{figure}[h]
 \centering
     \begin{minipage}{.5\textwidth}
\includegraphics[angle=0, width=3cm, height=1cm]{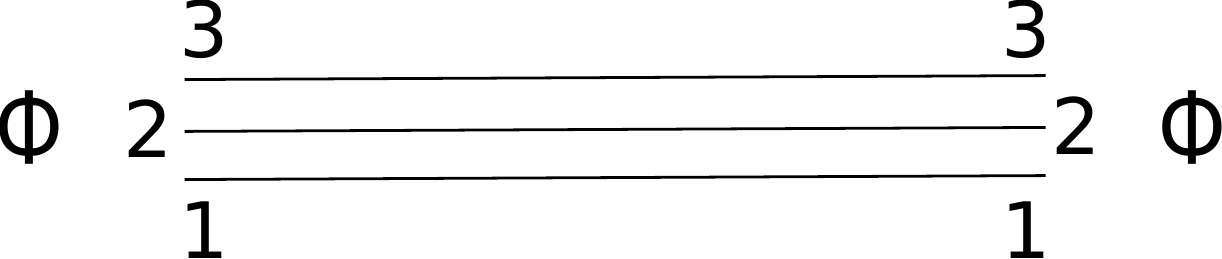} 
\caption{ {\small  The propagator.  }} 
\label{fig:prop}
\end{minipage}
\end{figure}

At the quantum level, the path integral is defined with a Gaussian field measure $d\nu_C(\phi)$ associated with a covariance or a propagator $C= (ZK+m)^{-1}$. The propagator is represented by  a stranded line, see Figure \ref{fig:prop}, which connects vertices of the form given by Figure \ref{fig:4vertex}.

\

Assuming the presence of a UV cutoff, i.e. for $|p_s|\leq M$, our mass and interaction terms are $O(2M+1)$ invariant, while the kinetic term breaks the $O(2M+1)$-invariance, just as in standard QFT the kinetic term breaks the locality property of the whole action. This type of kinetic term belongs naturally to our theory space as we will soon discover by straightforward expansion of our main RG equation. It also fits well with the cut-off prescription we will choose in the following for setting up our Wetterich FRG equation for GFTs.
%\
Furthermore, an almost identical model has been shown to have an interesting perturbative renormalisation flow in \cite{BenGeloun:2012pu} and it is indeed the simplest non-trivial model to be considered in $d=3$. The difference between the present model
and the model presented in that reference is that the tensor fields here 
are real rather than complex. As a result, the type of invariance
treated in \cite{BenGeloun:2012pu} was $U(2M+1)$ rather than $O(2M+1)$ as discussed in the  present work. With a kinetic term like \eqref{eq:3daction}, thus linear in \lq momenta\rq, the $U(2M+1)$ model is perturbatively
just-renormalisable at all orders. It has been also proved that this model is asymptotically free  \cite{BenGeloun:2012pu}. We will reveal, using a radically 
different tool as developed below, that the $O(2M+1)$ model is as well
asymptotically free. We will also argue that the perturbative 
calculation of the present model along the lines of the previous
reference will lead to the same conclusion (see Appendix \ref{app:pertur}). 

\

The interactions of the type \eqref{eq:3daction} are of the so-called {\it melonic} type, and combinatorially they correspond, when the field $\phi$ is represented by a triangle, to a subclass of triangulations of the sphere \cite{criticalTensor,Bonzom:2012hw}. They are the most dominant objects at large tensor size $M$ in the partition of function of colored tensor models (after an appropriate rescaling of action and coupling). As shown in \eqref{eq:3daction}, we shall use a single 
coupling $\lambda/3$ for the 3 possible connected interactions.
This specific combination of couplings is interesting on its own right and will lead to a unique coupling renormalised  equation for that combination.

 \

\section{The Wetterich equation for tensor models}
\label{sect:WE}

The functional renormalisation group (FRG) is an implementation of the the usual (Wilsonian) renormalisation group, well suited for practical calculations as well as for theoretical developments.
In essence, the idea is to trade the problem of evaluating the path integral for the problem of solving a functional differential equation, also known in general as FRG equation. While in principle this offers a perfectly equivalent (and equally difficult) reformulation of the problem, it allows for different approximations schemes. In particular, it allows to go beyond the standard perturbative approximation in dealing with the quantum interactions encoded in a field theory, and thus to explore in an efficient way its non-perturbative sector, notably its critical behaviour. It is mainly for this reason that the FRG has attracted a lot of attention over the last twenty years, leading to a great variety of results (see for example \cite{Kopietz:2010zz,Blaizot:2012fe,Gies:2006wv,Pawlowski:2010ht,Litim:2011cp} and references therein).

Our aim here is to generalize and apply the FRG methods to the Tensorial GFT models introduced in the previous section.
In doing so, we will partly be inspired by the similar developments for matrix models \cite{AstridTim, AstridTim2}, but we will also try to present the whole formalism in such a way to emphasize the field theoretical nature of the models. In other words, whereas vector and matrix models are usually thought of as $N$-component fields in zero dimensions, and a similar interpretation is possible for generic tensor models, our emphasis in dealing with TGFTs here is on their being field theories on specific $d\times D$-dimensional group manifolds (where $D$ is the dimension of the group $G$).

\subsection{FRGE formalism for tensor models }
\label{subsect:WE}

We adapt now the FRG formulation to tensor models described by \eqref{eq:actiond}.
Given the GFT perspective we mentioned above, the construction of the FRG formalism runs more or less along standard lines.

First we introduce a UV cut-off $M$ on the tensor modes\footnote{We use the labels \lq UV\rq \, and \lq IR\rq \,  for tensor modes according to the scales defined by the laplacian operator on the group manifold, or some other non-trivial kinetic term. That is, we stick to the mathematical role these modes play from the point of view of a field theory on the group manifold. We do not attach to these notion any spacetime or geometric interpretation. In fact, for TGFT models for quantum gravity, for which an interpretation in terms of simplicial geometry is possible for variables, quantum states, Feynman amplitudes, the immediate interpretation of these scales would run opposite to the label we use for them. In order to avoid confusion, and because the physical interpretation of the same models in terms of continuum spacetime and geometry is less straightforward, we avoid bringing such interpretative aspects into our presentation of the formalism and of our results.} and write the regularized partition function  as in \eqref{eq:partLam},
with source $J$, and free energy $W[J]$. From the free energy, one obtains the effective action by standard Legendre transform. The renormalisation group approach to quantum field theory rests on the equation governing the flow of effective actions at different scales, that is corresponding to a truncated partition function in which only modes between the UV cut-off and a certain IR cut-off are considered, as the IR cut-off is gradually removed. The critical behaviour of the system as the IR cut-off is removed will give indications about the phase diagram. In the same way, the possibility of complete removal of the UV cut-off, signifying that the quantum field theory is UV complete and thus well-defined, will be indicated by the presence of UV fixed points of the RG flow equation. 
The first step in obtaining the (functional) renormalisation group flow equation is thus to introduce a IR cut-off. 
An IR cut-off, which in our case will be parametrized by a pure number $N$, is implemented by adding to the action a term of the form
\be
 \label{DeltaS}
\Delta S_N [\phi]=\frac12 \Tr(\phi \cdot R_N  \cdot \phi ) 
=\frac12 \sum_{{\bf P},\tilde{{\bf P}}} \phi_{{\bf P}}\, R_{N}({\bf P};\tilde{{\bf P}}) \,\phi_{\tilde{{\bf P}}}\,. 
\ee
In particular, we take
\be \label{eq:R_N}
R_{N}({\bf P};\tilde{{\bf P}}) = N \bdel_{p_i,p_i'} R\Big(\fd\sum_{i=1}^d |p_i|/N\Big) \, ,
\ee
and impose on the profile function $R(z)$ the following conditions: positivity $R(z)\geq 0$ (otherwise \eqref{DeltaS} could emphasize rather than suppress modes), monotonicity $\frac{d}{dz} R(z) \leq 0$ (high modes should not be suppressed more than low modes), and lastly  $R(0)>0$ and $\lim_{z\to+\infty} R(z) = 0$ to exclude a constant (eventually zero) $R(z)$.
The last condition, together with the factor $N$ multiplying the profile function in \eqref{eq:R_N}, ensures that for $N\to 0$ the cutoff is removed. In the presence of a UV cutoff $M$ one usually adds also the condition that $\lim_{N\to M}R_{N}=\infty$, implying that in the limit that IR and UV cutoffs coincide no integration survives in the partition function.

Then, inserting the regulator $\Delta S_N$ in \eqref{eq:partLam},
we get a new regularized partition function of the form
\be\label{eq:partN}
\cZ_N[J] =
e^{W_N[J]}
 = \int_M d\phi\,  e^{- S[\phi] - \Delta S_N [\phi] + \Tr_2 (J \cdot \phi)} \,. 
\ee
The effective average action is defined as
\bea \label{EAA}
\Gamma_N[\varphi] = \sup_J \Big(\Tr_2 (J \cdot \varphi)  -  W_N (J) \Big)
- \Delta S_N [\varphi]\,.
\eea
Introducing the standard logarithmic scale $t = \ln N$, so that $\partial_{t} = N\partial_{N}$, deriving \eqref{eq:partN} with respect to $t$ and using \eqref{EAA} one arrives at
\beq
\partial_t \Gamma_N[\varphi] 
=
\frac12 \overline\Tr(\partial_t R_N \cdot [\Gamma^{(2)}_N + R_N]^{-1})
\label{eq:wetter}
\eeq
which is the Wetterich equation for an arbitrary rank-$d$
tensor model. $\overline{\Tr}$ is a ``super''-trace summing
over all momentum indices, $\Gamma^{(2)}_N=\delta^{(2)} \Gamma_N/(\delta \varphi_{{\bf P}} \delta \varphi_{\tilde{{\bf P}}} )$ is the second
derivative in the fields of the effective action. 
Note that the convolution between $\partial_t R_N$ and
the inverse of $\Gamma^{(2)}_N + R_N$ is performed in 
block tensor indices and it is matrix-like. The trace fully written reads
 $\sum_{{\bf P};{\bf P}'} \partial_t R_N({\bf P},{\bf P}') \cdot 
[\Gamma^{(2)}_N + R_N]^{-1} ({\bf P},{\bf P}')$.

An important feature of \eqref{eq:wetter} is the presence of $\partial_t R_N$ inside the super-trace.
Requiring that $R(z)$ and $\frac{d}{dz} R(z)$ go to zero fast enough for $z\to+\infty$ thus guarantees the UV finiteness of the super-trace, and as a consequence we can forget about the UV cutoff $M$.
Of course whenever we wish to talk in a meaningful way about the partition function we need to reintroduce the UV cutoff, but as already mentioned, the FRG perspective is to take \eqref{eq:wetter} as the fundamental equation replacing the path integral formulation.
Therefore, the standard approach is to try to study \eqref{eq:wetter} without any additional UV cutoff. 

However, as for any differential equation, we need initial conditions to construct a solution, 
\be
\Gamma_{N=M}[\varphi] = S[\varphi] \, ,
\ee
and these play essentially the role of UV scale and bare action.
The problem of constructing and solving the full continuum path integral thus translates into the problem of pushing the initial condition to $M\to\infty$, which usually requires the existence of a UV fixed point for the RG flow equation.

\subsection{Application: beta functions of a rank 3 model}
\label{subsect:Beta}

We wish now to apply the FRG formalism to the model described by the action $S$ 
in \eqref{simact} with constituents given by \eqref{eq:3daction}. 
As we just explained, we should in principle solve the full equation \eqref{eq:wetter} with $S$ as initial condition.
However, due to the functional and non-linear nature of the equation, this is a prohibitive task.
In particular, substituting the initial conditions into \eqref{eq:wetter}, i.e. evaluating the equation at $N=M$, and expanding the right-hand-side in powers of fields and momenta, one quickly realizes that an infinite series of new interactions not contained in the left-hand-side is being generated.
In order to obtain a closed and manageable system of equations a standard approximation strategy is that of choosing an ansatz for the effective action $\Gamma_N$, containing the initial condition $S$ as a special case, and truncating the expansion of the rhs of \eqref{eq:wetter} in such a way to only keep terms that match the original ansatz. The trustfulness of the outcome is then to be tested by extending the truncation, thus including more and more terms, and by studying the convergence of the results.

\

Here, to illustrate the formalism, we will consider the simplest possible truncation only, postponing to future work the investigation of larger theory spaces.
Therefore, we assume that the effective action 
takes the same form as $S$ and have
\bea
\Gamma_N(\varphi) = \frac{Z_N}{2}\Tr_2 (\varphi \cdot K \cdot \varphi)  +  \frac{m_N}{2} \Tr_2 (\varphi^2) + S^{\inter} \,,\qquad 
S'^{\inter}=  \frac{\lambda_N}{4} \Big( \Tr_{4;1} (\varphi^4) + \Sym (1 \to 2 \to 3)\Big) \,.
\eea

Specifying the regulator is the next task, for which we choose to introduce a Litim-like regulator \cite{Litim:2001up}:
\be \label{eq:ZR_N}
R_{N}(\{p_i\};\{\tilde{p}_i\})
=  Z_N\, \bdel_{p_i,\tilde{p}_i} \,
\Big(N - \ft\sum_{i=1}^3 |p_i|\Big) \,\Theta(N - \ft\sum_{i=1}^3|p_i|)\,. 
\ee
Note that 
$R_{N}$ is symmetric in the following sense: 
$R_{N}(\{p_i\};\{\tilde{p}_i\}) = R_{N}(\{\tilde{p}_i\};\{p_i\}) $.
In comparison to \eqref{eq:R_N} we have included in the regulator also a wave function renormalisation $Z_N$.
One can check that this regulator kernel obeys the cutoff conditions, provided that $Z_N >0$ for all $N$ and $\lim_{N\to 0}N/Z_N=0$ (because we want to be able to remove the cutoff even for $p_i=0$).

A small calculation shows that
\beq
\partial_t R_{N} (\{p_i\};\{ \tilde{p}_i\}) =\bdel_{p_i,\tilde{p}_i} \Big\{ \,\partial_t Z_N (N - \ft\sum_{i=1}^3 |p_i|)   \,+ Z_N N \Big\} \Theta(N - \ft\sum_{i=1}^3 |p_i|)  \,,
\eeq
which is again a symmetric kernel. 

Evaluating $\Gamma_N^{(2)}$, one gets 
\bea
&&
\Gamma_N^{(2)}(\{p_i\};\{p_i'\}) = \Big( Z_N\ft\sum_{i}|p_i|  +  m_N \Big) \bdel_{p_i,p_i'}\, 
 + F_N(\{p_i\};\{p_i'\})\,,\crcr
&& 
F_N(\{p_i\};\{p_i'\}) = 
\lambda_N\Big[ \sum_{m_2,m_3}\varphi_{p_1,m_2,m_3} 
\varphi_{p_1',m_2,m_3} \delta_{p_2,p_2'}\delta_{p_3,p_3'}+
\sum_{m_1} \varphi_{m_1,p_2,p_3}\varphi_{m_1,p'_2,p'_3}\delta_{p_1,p_1'}
+   \varphi_{p'_1,p_2,p_3}\varphi_{p_1,p'_2,p'_3}  \cr\cr
&+& \Sym(1\to 2 \to 3) \Big]   \,. 
\eea
Note that there is a symmetry factor of 4 related to the
fact that we can first choose any leg of the vertex
to perform the initial derivation.  

The next stage is to perform the expansion
of the r.h.s. of the  Wetterich equation \eqref{eq:wetter}. Let us denote
\bea
P_N(\{p_i\};\{p_i'\}) &=& R_N(\{p_i\};\{p_i'\}) +  (Z_N\ft\sum_i |p_i| + m_N)  \bdel_{p_i,p_i'} 
 \,.
\eea
Expanding \eqref{eq:wetter}, we obtain 
\bea
 \partial_t \Gamma_n  &=& \frac12 \overline\Tr \Big(\partial_t R_N \cdot 
 [\Gamma^{(2)}_N + R_N ]^{-1}\Big)  \cr\cr
 &=& 
 \frac12 \Big[ 
 \overline\Tr\Big(\partial_t R_N \cdot
 P_N^{-1}\Big )+ 
 \sum_{n=1}^{\infty} (-1)^n \overline\Tr\Big(\partial_t R_N \cdot
 P_N^{-1}\cdot[ F_{N} (P_N)^{-1}]^{n}\Big)\Big] \,.
 \eea

After a lengthy calculation (the most relevant aspects of this computation are reported in Appendix \ref{app:wett}), we arrive at
the following system of ordinary differential equations: 

\

$\bullet$ for the wave function $ Z_N$,

\beq \label{etafull}
\boxed{\partial_t Z_N =\frac{54 \lambda_N Z_N N (2
   N+1)}{3 (Z_N N+ m_N)^2- 2 \lambda_N ( 27 N^2+ 18 N +5)}}
\eeq

$\bullet$ for the mass $m_N$, 

\beq \label{mfull}
 \boxed{  \partial_t m_N  = \frac{-\lambda_N\, N}{(Z_N N+ m_N)^2}
\Big(
  (18N^2 +9N + 4 ) \partial_t Z_N+  ( 54 N^2+36N+9) Z_N
\Big)}
\eeq

$\bullet$ for the coupling constant $\lambda_N$, 

\beq \label{lambdafull}
 \boxed{\partial_t \lambda_N  = \frac{2 \lambda_N^2  N}{(Z_N N+m_N)^3}\Big( \ft  \,(18N^2+45N+  37 ) \partial_t Z_N
+18 (N+1)^2 Z_N\Big)} \,. 
\eeq

This is the complete set of RG flow equations for our model, in the chosen truncation.

We emphasize that the calculation is performed at fixed
$N$ without any approximation of the equation or any simplification for the fields (for instance we do not assume they are proportional to delta functions, as in \cite{AstridTim}). 
In addition to this technical aspect and to the difference in perspective noted in the introduction, we notice here a few more technical differences with the work of Eichhorn and Koslowski
\cite{AstridTim, AstridTim2}, dealing with a
FRG treatment of matrix models. First, the very extension to tensors implies a richer set of invariant monomials that can be considered as interaction terms, thus a richer theory space. Second, 
we will include in our truncation a non-trivial kinetic term, with an associated $O(N)$-breaking, as similar terms would anyway be generated by the RG flow.

We should also point out that in the summations reported in Appendix \ref{app:wett} the variable $N$ is treated as an integer. For consistency, we should either replace the derivatives in  \eqref{mfull} and \eqref{lambdafull} with finite difference operators, if insisting on integer $N$, or we should substitute $N$ with the floor function $\lfloor N \rfloor$ in the polynomials arising from the summations, if treating $N$ as a continuous parameter. We will take the latter point of view, but as we will mostly be interested in the large- or small-$N$ limits, the staircase nature of the beta functions will not be important for our analysis.\footnote{The staircase nature of the functional traces is obviously due to the use of a non-smooth cutoff (containing a step function) in combination with a discrete spectrum of modes. A situation of this type has been encountered and discussed for example in works on asymptotic safety using a spherical background metric \cite{Benedetti:2012dx,Demmel:2014sga}. This is part of the scheme dependence associated to the cutoff choice, but we expect that the main qualitative features of our analysis would remain unaltered when using a smooth cutoff.}

\

%%%%%%%%%%%%%%%%%%
\section{Flow equations, fixed points and critical exponents}
%%%%%%%%%%%%%%%%%%
\label{sect:Flows}

%%%%%%%%%%%%%%%%%%
\subsection{$N$-dependent beta functions and scaling of the couplings}
%%%%%%%%%%%%%%%%%%
\label{subsect:discNdep}

In the previous section, we have obtained the RG flow equations \eqref{mfull} and \eqref{lambdafull} for our example model, which we now want to analyze.
The standard first step in any analysis of flow equations is to rescale the couplings with the wave function renormalisation and the running scale in such a way to obtain a system of ordinary differential equations for the couplings with no explicit dependence on the running scale and with no dependence at all on $Z_N$.
The fact that it is generically possible to eliminate  $Z_N$ from the flow equations is a consequence of the reparametrization invariance of the RG flow, for which the presence of $Z_N$ in the cutoff \eqref{eq:ZR_N} is essential \cite{Berges:2000ew}. As usual, this is achieved by defining the anomalous dimension $\eta \equiv \partial_t \ln Z_N$ and the rescaled couplings $m_N \to Z_N m_N$, $\lambda_N\to Z_N^2 \lambda_N$, leading to
\be  \label{etaZ}
\eta =\frac{54 \lambda_N N (2   N+1)}{3 (N+ m_N)^2-2 \lambda_N ( 27 N^2+ 18 N +5)} \, ,
\ee
\be \label{mZ}
  \partial_t m_N  = \frac{-\lambda_N N}{(N+m_N)^2}
\Big[
\eta \, (18N^2 +9N +4 )+ (54 N^2+36N+9) 
\Big] - \eta\, m_N \, ,
\ee
\be  \label{lambdaZ}
\partial_t \lambda_N  = \frac{2 \lambda_N^2  N}{(N+m_N)^3}\Big( \ft \eta \,(18N^2+45N+  25 )
+  18 (  N+1)^2 \Big) -2\, \eta\, \lambda_N \, .
\ee

On the other hand, removing any explicit dependence on the running scale, and therefore obtaining an autonomous system for the couplings, typically requires the absence of fixed external scales. In such case, the statement that we can obtain an autonomous system is equivalent to the statement that we can switch to dimensionless couplings, since for obvious dimensional reasons, the beta functions for the dimensionless couplings cannot depend on the running scale unless we can form a dimensionless ratio between this and some external scale. Examples of non-autonomous systems due to the presence of external scales are known in the literature, for example  quantum field theory at finite temperature \cite{Berges:2000ew}, or on a curved \cite{Benedetti:2014gja} or non-commutative spacetime \cite{Gurau:2009ni}.

In the present case, due to the polynomials in $N$ that appear in the beta functions, it is impossible to find a simple rescaling of the couplings with some power of $N$ that would lead to an autonomous system\footnote{In fact, it is impossible even with a more generic  rescaling $\lambda_N\to f(N) \lambda_N$.}.
We can ask ourselves why do we have such a situation here. Obviously $N$ is a pure number, hence it does not depend on any units, and there is no reason why it should not appear explicitly in the beta functions. This is the natural situation in matrix or tensor models, therefore it might seem superfluous to ask for a deeper explanation. However, the field-theoretic perspective on Tensorial GFTs provides us with an explanation that highlights the similarity to previous examples of non-autonomous RG systems.
In fact, our rank-3 tensorial model is a field theory on $S^1\times S^1 \times S^1$ (with an unusual non-local interaction), where we have fixed the radius of $S^1$ to 1, or, in other words, where we have expressed every dimensionful quantity, including the eigenvalues of the Laplacian, in units of the radius $L$. It is such hidden radius that provides the fixed external scale, in terms of which we can write $N=kL$, being $k$ the standard (dimensionful) RG running scale, thus giving us an interpretation for the non-autonomous system.

The non-autonomous nature of the system complicates slightly our analysis, as  the ``force field'' is now ``time''-dependent, and fixed points are unlikely\footnote{For a non-autonomous dynamical system $\partial_t g_i = \beta_i(g,t)$, the zeros of the beta functions, $\beta_i(g^*,t)=0$, are generically time-dependent, $g_i^*=g_i^*(t)$ (unless a $t$-independent part factors out, $\beta_i(g,t)=f(g,t) h(g)$ and $h(g^*)=0$). As a consequence, $\partial_t g^*_i\neq 0$ and such zeros are not solutions of the flow equations. It can happen, however, that if the limit for $t\to\pm\infty$ of $g_i^*(t)$ exists, certain (possibly trivial) subsets of trajectories are attracted to it in the same limit, as we will discuss below.}.

We can of course integrate numerically the system \eqref{mZ}-\eqref{lambdaZ} as it is, and plot representative trajectories for $m_N$ or $\lambda_N$ as function of $N$, or in a 3D-space $\{m_N,\lambda_N,N\}$.
However, the analysis becomes more transparent if we first separately study the two opposite regimes  $N\gg 1$ and $N\ll1$.
In the limit of large or small $N$ the polynomials appearing in the beta functions can be approximated by the respective leading monomials and as a result it becomes possible to rescale the couplings in such a way to obtain an autonomous system.

%%%%%%%%%%%%%%%%%%
\subsection{Large-$N$ limit}
%%%%%%%%%%%%%%%%%%
\label{subsect:LargeN}

We first study the large-$N$ limit, which in light of the previous discussion, we can interpret also as the limit of $k\gg L^{-1}$. 
It turns out that in this limit we only need to define a rescaled mass via
\be \label{mbar}
m_N = N \bar m_N \, 
\ee
in order to obtain the following autonomous system:
\be  \label{etaLargeN}
\eta =\frac{36 \lambda_N }{  (1+\bar m_N)^2 - 18 \lambda_N} \, ,
\ee
\beq \label{mLargeN}
  \partial_t \bar m_N  = -18\frac{\lambda_N}{(1+\bar m_N)^2}
(\eta + 3) - (1+\eta)\, \bar m_N \, ,
\ee
\be  \label{lambdaLargeN}
\partial_t \lambda_N  = 12\frac{ \lambda_N^2}{(1+\bar m_N)^3}(  \eta + 3) -2\, \eta\, \lambda_N \, .
\ee

Having now an autonomous system, we can apply the standard analysis. First of all we look for fixed points, i.e. for zeros of the beta functions. We find the usual Gaussian fixed point (GFP) at $\bar m_N=\lambda_N=0$, plus two non-Gaussian fixed points (NGFPs) at
\be \label{largeNGFP}
\bar m^*_\pm = -\frac{14 \pm \sqrt{7}}{21} \simeq  \left\{ \begin{aligned} -0.7926 \\ -0.5407 \end{aligned} \right. \, , \;\;\;
 \lambda^*_\pm = - \frac{\bar m^*_\pm}{189}\simeq  10^{-3}\times \left\{ \begin{aligned} 4.193 \\ 2.861 \end{aligned} \right.\, .
\ee

Next, we can study the linear stability properties around the fixed points, i.e. compute critical exponents and eigen-perturbations of the flow\footnote{
As a reminder, the general procedure goes as follows. Given a dynamical system $\partial_t g_i = \beta_i(g)$, and a fixed point $g^*_i$ such that $\beta_i(g^*)=0$ for all $i$, then we can expand the system to linear order
\be
\partial_t g_i = \sum_j B_{ij} (g_j-g^*_j)\, ,
\ee
where $B_{ij}\equiv \partial \beta_i/\partial g_j {}_{|g=g^*}$. The linearized system can be easily solved by diagonalizing the stability matrix $B_{ij}$. Defining the critical exponents $\theta_I$ as minus its eigenvalues, $\sum_j B_{ij} V_j^I = - \theta_I V_i^I$, we obtain
\be
g_i(t) = g_i^* + \sum_I C_I V^I_i e^{-\theta_I t}\, ,
\ee
where the $C_I$ are integration constants. Relevant eigen-perturbations $V^I$ are the ones which are important in the IR (i.e. for $t\to-\infty$), hence they correspond to positive critical exponents $\theta_I>0$. Irrelevant eigen-perturbations are those with $\theta_I<0$, and the marginal ones those with vanishing exponent.
}.
For the GFP, the critical exponents correspond to the canonical scaling dimension of the couplings, i.e. in this case $\theta^{(0)}_1=1$ and $\theta^{(0)}_2=0$, with eigenvectors $V^1=(1,\, 0)$ and $V^2=(-54,\, 1)$ respectively. Here by canonical scaling dimension we mean the power of $N$ in the rescaling that leads to an autonomous system, which in this case is \eqref{mbar} for the mass and no rescaling for $\lambda_N$. We discuss a different point of view on scaling dimensions in Appendix \ref{App:dim}.
The scaling dimension of $\lambda_N$ being zero, the quartic interaction (or more precisely $-54 \Tr_{2}(\phi^2)+\Tr_{4}(\phi^4)$) is a marginal perturbation of the GFP. In order to understand the true nature of the perturbation we have to expand the beta functions beyond linear order. At quadratic order in the couplings we have
\be \label{betam2}
\partial_t \bar m_N \simeq - \bar m_N - 54 \lambda_N +72 \bar m_N \lambda_N -648 \lambda_N^2 \, ,
\ee
\be \label{betal2}
\partial_t \lambda_N \simeq - 36 \lambda_N^2 \, .
\ee
The equation for $\lambda_N$ tells us  immediately that we have a marginally relevant interaction, i.e. we have asymptotic freedom\footnote{Equation \eqref{betal2} is easily solved for
\be
\lambda_N = \frac{1}{A+36 \ln N}\, ,
\ee
where $A$ is an integration constant. We recognise the standard logarithmic approach to asymptotic freedom ($\lambda_N\to 0$) for $N\to +\infty$.
Plugging this solution into \eqref{betam2} we can solve also this other equation exactly. The complete solution is a bit lengthy and not very important here, but its asymptotic behavior gives
\be
m_N \sim -\frac{1}{ \ln N}\left( \frac{3}{2} +O\left( \frac{1}{\ln N}\right) \right) + \frac{1}{N} \left( B (\ln N)^2 + O\left(\ln N\right) \right)\, ,
\ee
where $B$ is another integration constant. }. This result is in agreement with the perturbative calculation in \cite{BenGeloun:2012pu}, see Appendix \ref{app:pertur}.

For the NGFP at $\{\bar m^*_-,\lambda^*_-\}$ the critical exponents are %
\be
\theta^{(-)}_\pm = \frac{1}{4} \left( 17\sqrt{7}-47 \pm \sqrt{2776-1038\sqrt{7}} \right) \simeq  \left\{ \begin{aligned} 0.8571   \\ -1.868\end{aligned} \right.  \, ,
\ee
and therefore we have one relevant and one irrelevant eigen-perturbation.

\

One should notice also that due to the nonlinear, rational, nature of the beta functions, we also have singularities at $\bar m=-1$ and at $\lambda=(1+\bar m)^2/18$. The appearance of such type of singularities is a generic feature of truncated FRG flows, and generally it is understood as an artifact of the truncation, or of the parametrization, as we discuss below. 
Therefore we do not trust the validity of the truncation in the vicinity of the singularity. We are also generally interested in flow trajectories that are connected to the free theory, hence we will not take into consideration the region which is disconnected, because of the singularity, from the origin. This region includes the second NGFP located at $\{\bar m^*_+,\lambda^*_+\}$, which therefore we will not discuss further, and from here on by NGFP we will mean the one at $\{\bar m^*_-,\lambda^*_-\}$.
%%%%%%%%%%%%
\begin{figure}[ht]
\begin{center}
\includegraphics[width=8cm]{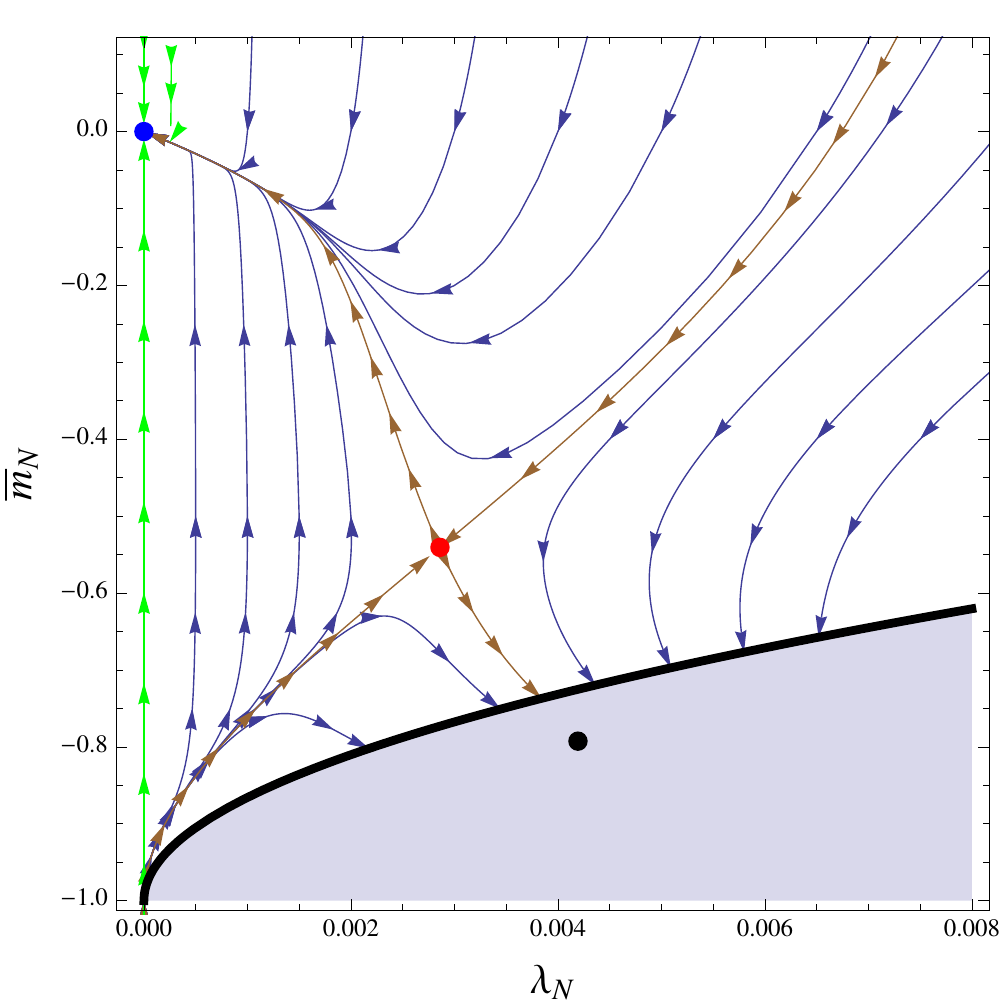} 
\caption{
The flow diagram at large $N$. The blue dot is the GFP, while the red and black one are  $\{\bar m^*_-,\lambda^*_-\}$ and  $\{\bar m^*_+,\lambda^*_+\}$ respectively. The shaded region is disconnected from the GFP because of the singularity (thick black line). Ordinary trajectories are in blue, while the eigen-perturbations for the GFP are in green and those for the NGFP are in brown. Arrows point towards the UV, i.e. growing $N$.}
\label{Fig:flow-largeN}
\end{center}
\end{figure}
%%%%%%%%%%%%

Finally, we can numerically integrate the large-$N$ flow, and the resulting phase diagram is shown in Figure~\ref{Fig:flow-largeN}.
Strikingly, the picture that emerges from our analysis is very similar to the standard scalar case in three dimensions. We have a GFP with two relevant directions, and a NGFP with only one relevant direction. 
It is however important to remark the differences. Whereas one of our two relevant directions at the GFP is only marginally relevant, in standard scalar field theory in three dimensions the two relevant  eigen-perturbations at the GFP are strictly relevant (in addition there is a marginally irrelevant one, the $\phi^6$ interaction). 
Even more strikingly, if we consider, always in three dimensions, a scalar field theory with a standard $\phi^4$ interaction, but a non-standard kinetic term which is linear in momentum (to mimic our model) then  the quartic interaction would be irrelevant.\footnote{Such model belongs to the class of long-range interaction models recently studied in \cite{Defenu:2014bea} by FGR methods.}
Therefore we conclude that we should attribute the different behavior entirely to the non-standard non-local interaction of our model. This is one further example, beside the recent results on perturbative renormalizability of TGFT models, of how the peculiar combinatorial structure of TGFT interactions affects their behaviour as QFTs, defying what we would expect from conventional QFTs.

\

Besides the marginal nature of the interaction, at a qualitative level,  our NGFP is very reminiscent of the Wilson-Fisher fixed point, and just as that, it appears as an IR fixed point for the continuum theory defined by the GFP.
We can then interpret as usual the region above the irrelevant trajectory at the NGFP as the symmetric phase (both $\lambda_N$ and $m_N$ are positive in the IR), and the one below as the broken or condensed phase (in the IR we have positive $\lambda_N$ and negative $m_N$). 

In the broken phase, this picture is again limited by the presence of singularities. As we already mentioned, such singularities occur also in the standard scalar field case, and it is well understood that this is due to the choice of parametrization of the potential. In fact, it turns out that in the broken phase it is better to write the potential as $V(\phi) = \lambda (\phi^2- \phi_0^2)^2$, and study the running of the coupling $\lambda$ and of the minimum $\phi_0$. In terms of such variables there is no singularity for $\phi_0^2\geq 0$.
In the tensorial case we can presume that the same would be true, but due to the non-local nature of the interaction term, the minimization of the potential is non-trivial.
In any case, we should remember that the phase transition associated to the diagram in  Figure~\ref{Fig:flow-largeN} is only going to be a true phase transition (in a strict sense) if it survives a small-$N$ analysis.
In fact, the analysis above was done for large $N$, and as a consequence, the plot in  Figure~\ref{Fig:flow-largeN} should be interpreted as approximately valid from some large value of $N$ and above. In other words, for an initial condition at large $N$, represented by a point  in  Figure~\ref{Fig:flow-largeN}, the trajectory passing through it is trustable in its forward part (i.e. following the direction of the arrows), but in the opposite direction it has a validity which is limited to the condition $N\gg 1$, i.e. we cannot follow the flow backward indefinitely. 
Since reaching a fixed point always takes an infinite amount of time, we conclude that even along the irrelevant eigen-perturbations at the NGFP we do not really reach the fixed point in the IR, and thus its interpretation as an IR fixed point should only be taken in the sense that it is an IR attractive point for the flow as long as we are not too much in the IR.
Going backward, towards small $N$ the approximation breaks down and we need to get back to the full original non-autonomous system  \eqref{mZ}-\eqref{lambdaZ}. After a transient region, which can only be studied by numerically integrating  \eqref{mZ}-\eqref{lambdaZ}, we must end up in the small $N$ regime, where again we can approximate the $N$-polynomials with their lowest order monomials and perform a standard analysis, as we are going to discuss in the following subsection.
The above warning about the limits of validity of the phase diagram obtained for large-N assumes of course a fixed external scale $L$ for the group manifold. On the other hand, one could expect the same phase diagram to be valid in its entirety, and to represent the whole RG flow equations, in the limit $L\to\infty$, i.e. in the limit of very large domain manifold for the TGFT fields. We will discuss this point as well in the following subsection.

%%%%%%%%%%%%%%%%%%
\subsection{Small-$N$ limit}
%%%%%%%%%%%%%%%%%%
\label{subsect:smallN}

Due to the step function in the cutoff, the small-$N$ behavior is drastically non-smooth.
As soon as $N<1/3$, the only surviving modes in the functional traces are those with $p_i=0$, i.e. the zero modes of the GFT fields.
Using the results in Appendix \ref{app:wett} we find easily that in this regime the anomalous dimension vanishes identically, and
\be
  \partial_t m_N  = -9 \frac{\lambda_N N}{(N+m_N)^2}\, ,
\ee
\be
\partial_t \lambda_N  = 36 \frac{ \lambda_N^2  N}{(N+m_N)^3}  \, .
\ee

The only consistent rescaling leading to an autonomous system is the one given by \eqref{mbar} together with
\be \label{lambdabar}
\lambda_N = N^2 \bar \lambda_N 
\ee
from which we obtain
\be \label{mSmallN}
 \partial_t \bar m_N=  -\bar m_N  -9 \frac{\bar\lambda_N\, }{(1+\bar m_N)^2}    \, ,
\ee
\be \label{lambdaSmallN}
\partial_t\bar\lambda_{N}  =  - 2    \bar\lambda_{N}  + 36\, \frac{\bar\lambda_N^2}{(1+\bar m_N)^3} \, .
\ee
The same result is obtained directly from  \eqref{mZ}-\eqref{lambdaZ} in the $N\to 0$ limit, forgetting about the staircase nature of the spectral sums.

The system admits again a GFP, but being this a UV fixed point, the previous large-$N$ approximation is the one appropriate to its study.
More interestingly, we find a NGFP at
\be \label{smallNGFP}
\bar m^* = -\frac{1}{3} \, , \;\;\; \bar \lambda^* = \frac{4}{243} \, .
\ee
In this case we have no other NGFP, and like for the  large-$N$ NGFP located at $\{\bar m^*_-,\lambda^*_-\}$, there is no singularity separating it from the GFP (now we only have the singularity at $\bar m=-1$).
Its critical exponents are
\be
\theta^{*}_+ = 1\, , \;\;\; \theta^{*}_- = -3  \, ,
\ee
that is, we find again one relevant and one irrelevant direction.
Interestingly, however, the critical exponents are integer numbers, a property that usually characterises the Gaussian fixed points.

\

\begin{figure}[ht]
\begin{center}
\includegraphics[width=8cm]{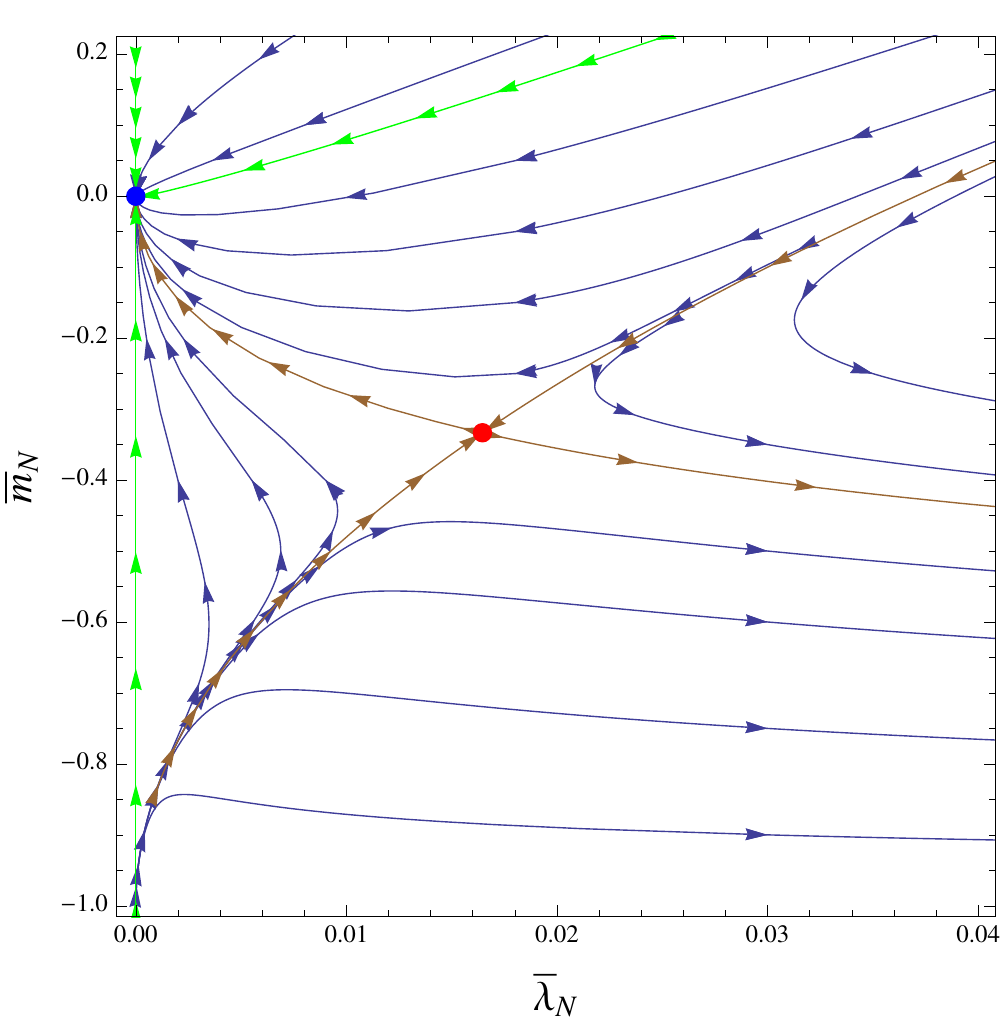} 
\caption{
The flow diagram at small $N$. The blue dot is the GFP, while the red one is the NGFP at $\{\bar m^*,\bar\lambda^*\}$. Ordinary trajectories are in blue, while the eigen-perturbations for the GFP are in green and those for the NGFP are in brown. Arrows point towards the UV, i.e. growing $N$.}
\label{Fig:flow-smallN}
\end{center}
\end{figure}

The phase diagram resulting form the numerical integrations of \eqref{mSmallN}-\eqref{lambdaSmallN} is shown in Figure~\ref{Fig:flow-smallN}.
%%%%%%%%%%%%
%%%%%%%%%%%%
The qualitative characteristics of the diagram are similar to those found at large-$N$ in the region connected to the GFP, and we feel encouraged to interpret the NGFP \eqref{smallNGFP} as the small-$N$ counterpart to the large-$N$ NGFP \eqref{largeNGFP} located at $\{\bar m^*_-,\lambda^*_-\}$.

\

However, we should stress that such NGFPs were obtained from different rescaling of $\lambda$, and going back to the original coupling via \eqref{lambdabar}, we notice that for $N\to 0$ the NGFP \eqref{smallNGFP} corresponds to $\lambda^* = 0$, while the one in \eqref{largeNGFP} was at $\lambda^* \neq 0$.

This observation could also explain the integer critical exponents. Even though $\bar m_N$ and $\bar\lambda_N$ have a nontrivial fixed point, the scaling \eqref{mbar} and \eqref{lambdabar} implies that at such fixed point the renormalised mass and the renormalised coupling (i.e. their value in the limit $N\to 0$) are zero.
Once again, modulo an exchange in the scaling dimensions of mass and coupling, the same conclusion can be reached for the standard Wilson-Fisher fixed point in three dimensions. However, in such a case we can easily study higher-order truncations, and find that also the coupling $g_6$ of the $\phi^6$ interaction reaches a fixed point, and being $g_6$ dimensionless in $d=3$, it remains finite also as we remove the IR cutoff. That the Wilson-Fisher fixed point theory is truly an interacting one, can also be inferred more reliably from the local potential approximation or the next orders in the derivative expansion \cite{Bagnuls:2000ae}.
In the Tensorial GFT case, on the other hand, we are not able to do a full local potential approximation, but from our truncation we can easily guess that the IR scaling dimension for the coupling of a general interaction is \eqref{gn-dim} with $\alpha=0$, and hence all such couplings would flow to zero at an IR fixed point. The non-trivial fixed point is really a trivial one in disguise.
We also notice that such scaling dimensions for the couplings are the one we would get for standard couplings in zero dimensions, where we expect no phase transition and no non-trivial fixed point.

Figure~\ref{Fig:flow-smallN} might seem to contradict such expectation at first, but in fact a similar flow diagram is found by analytically continuing the usual beta equations to $d=0$ (which in fact have the same structure as \eqref{mSmallN}-\eqref{lambdaSmallN}). The explanation of the apparent paradox is again found by remembering that in the broken phase we should better use a more appropriate truncation, such as $V(\phi) = \lambda (\phi^2- \phi_0^2)^2$. Then one finds that in zero dimensions the non-trivial fixed point is IR attractive for both $\lambda$ and $\phi_0^2$, and it lies at $\phi_0^2<0$, meaning that actually there is always symmetry restoration in the deep IR. Although we cannot at the moment repeat this analysis from scratch in the Tensorial GFT case, the similarity of the equations in the symmetric case, together with the scaling argument, give us confidence that the same is true here.

The fact that the zero modes surviving in the deep IR lead to an effective zero-dimensional theory is very reminiscent of what observed in \cite{Benedetti:2014gja} for  scalar field theory on a spherical background. Just like in that case, also in our case we can trace back the origin of such phenomenon to the compactness of the background space, which  in \cite{Benedetti:2014gja} was $S^d$, while here is $(S^1)^d\simeq T^d$. 

\

All in all, for a quantum field theory on a compact space we would not expect a phase transition, on general grounds, and our results seem to confirm this in the Tensorial GFT case as well, and the apparent NGFP is most likely an artefact of the parametrisation chosen.

At the same time, this suggests that a phase transition could be found also for our GFT model (and other GFTs as well) in a ``flat" regime, corresponding to the limit of very large radius of the compact manifold $L\to\infty$, in which the group elements are well approximated by their algebra elements\footnote{This is similar to what happens in ordinary field theories that, while initially defined in a box (and often with periodic boundary conditions), are then studied in the thermodynamic limit (corresponding to infinite box) to unravel their thermodynamic, large scale properties, in particular their phase diagram. This is true even when the system is, physically, a finite one.}.

In this regime, due to the way the radius of the compact manifold enters the theory, we expect the RG flow to match the one we obtain in the large-$N$ approximation, and the corresponding phase diagram, which we have analysed in the previous subsection, describes the whole phase structure of the system. Indeed, from this phase diagram one can see evidence of a phase transition from a symmetric to a broken or condensed phase, as we have explained. We note, however, that  a proper, detailed scaling of the model, and the proper definition of a thermodynamic limit for GFTs in general, is tricky due to the non-local structure of the interactions and various other subtleties related to the quantum gravity interpretation of these models, and we leave it for future work. 

Finally, the above results suggest that a true phase transition to a broken phase, without passing through any further approximation, is only to be expected for GFTs on non-compact manifolds, like Lorentzian quantum gravity models. This is in itself interesting, as it may give an independent argument, entirely based on RG considerations, to believe that some form of causality  is needed to have a proper continuum geometric phase\footnote{In the GFT/spin foam case, the use of Lorentz group implies the presence of a local light cone structure in each cellular complex corresponding to a Feynman diagram of the GFT model considered, but no global foliations or the like, as it is the case in causal dynamical triangulations.}.

%%%%%%%%%%%%%%%%%%
\subsection{General $N$}
%%%%%%%%%%%%%%%%%%

We should stress again that the NGFPs discussed in the two previous subsections are not proper fixed points, as they are not solutions to the full system  \eqref{mZ}-\eqref{lambdaZ}. 
The closest object to a fixed point would be a complete trajectory linking $\{\bar m^*,\bar\lambda^*\}$ to $\{\bar m^*_-,\lambda^*_-\}$, but finding such a trajectory is a highly non-trivial problem: we have no analytical proof, and being both saddle points, finding it numerically  is a very difficult task.

\begin{figure}[ht]
\begin{center}
\includegraphics[width=5.5cm]{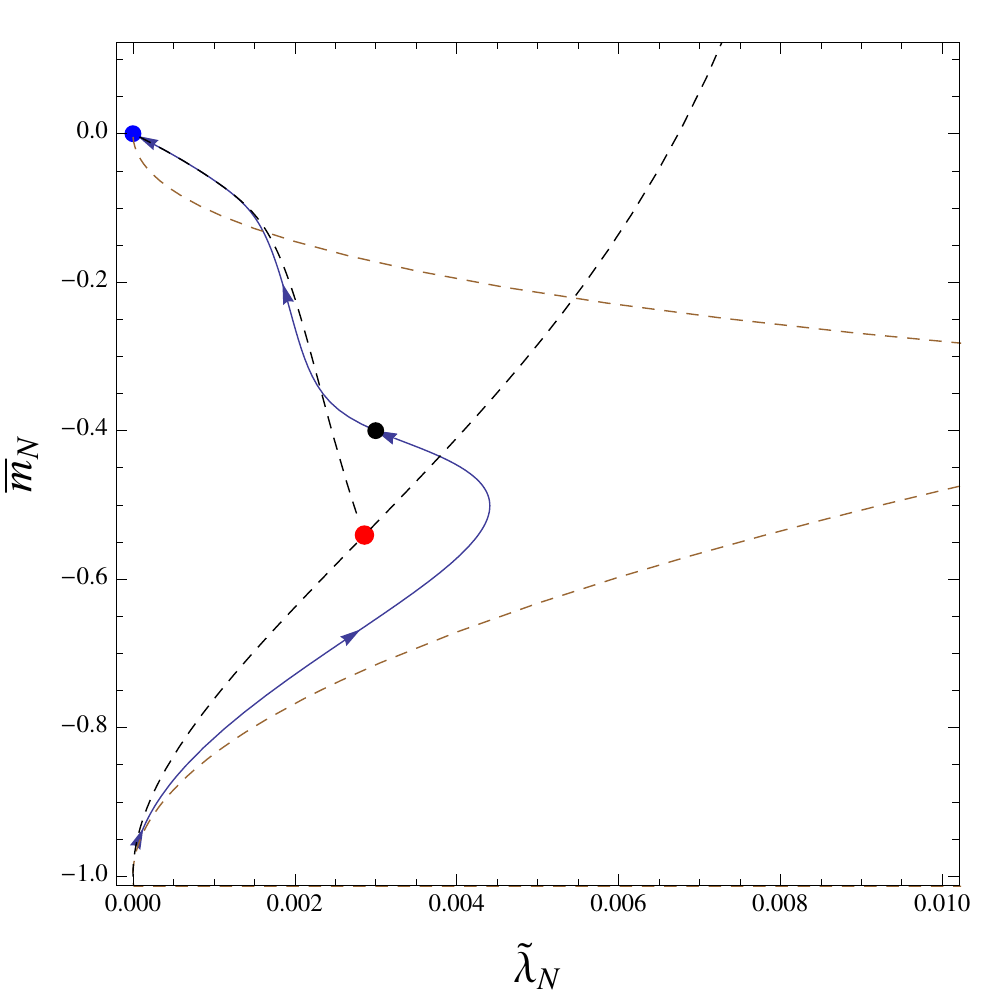} \hspace{.1cm} \includegraphics[width=5.5cm]{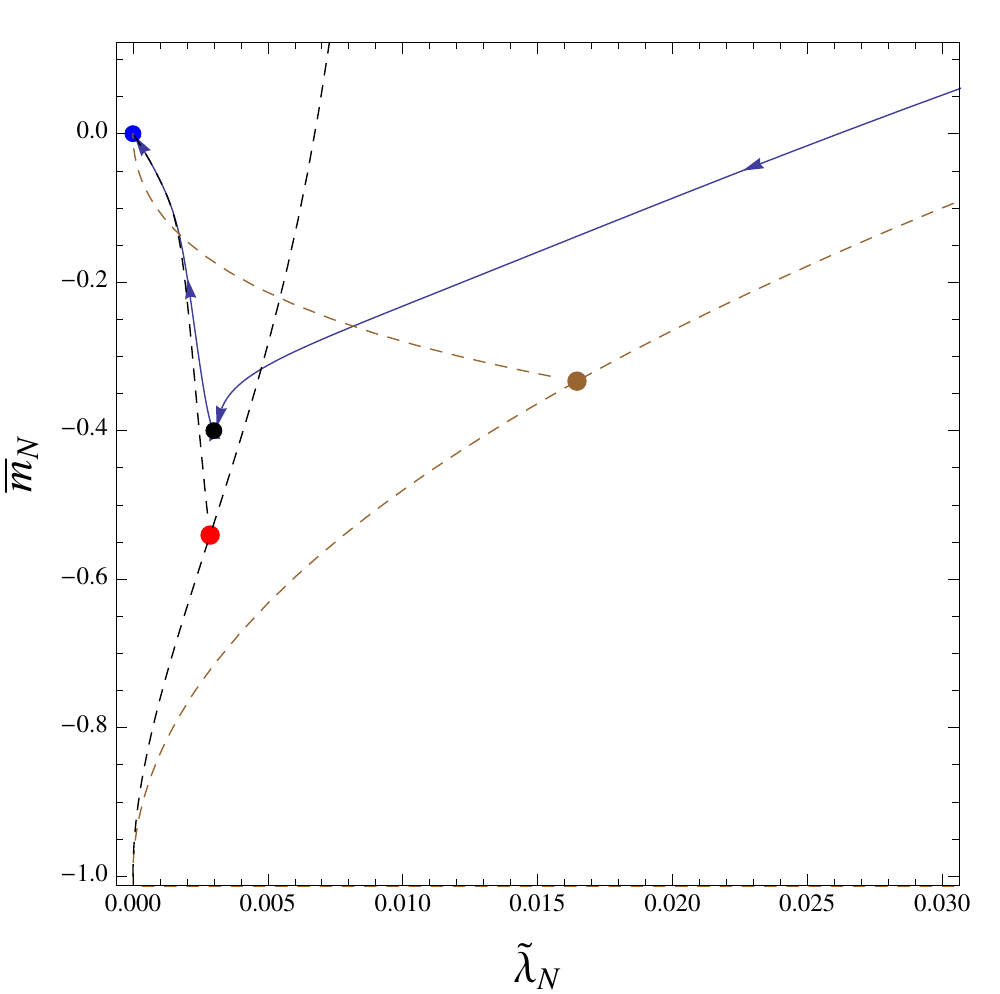}  \hspace{.1cm} \includegraphics[width=5.5cm]{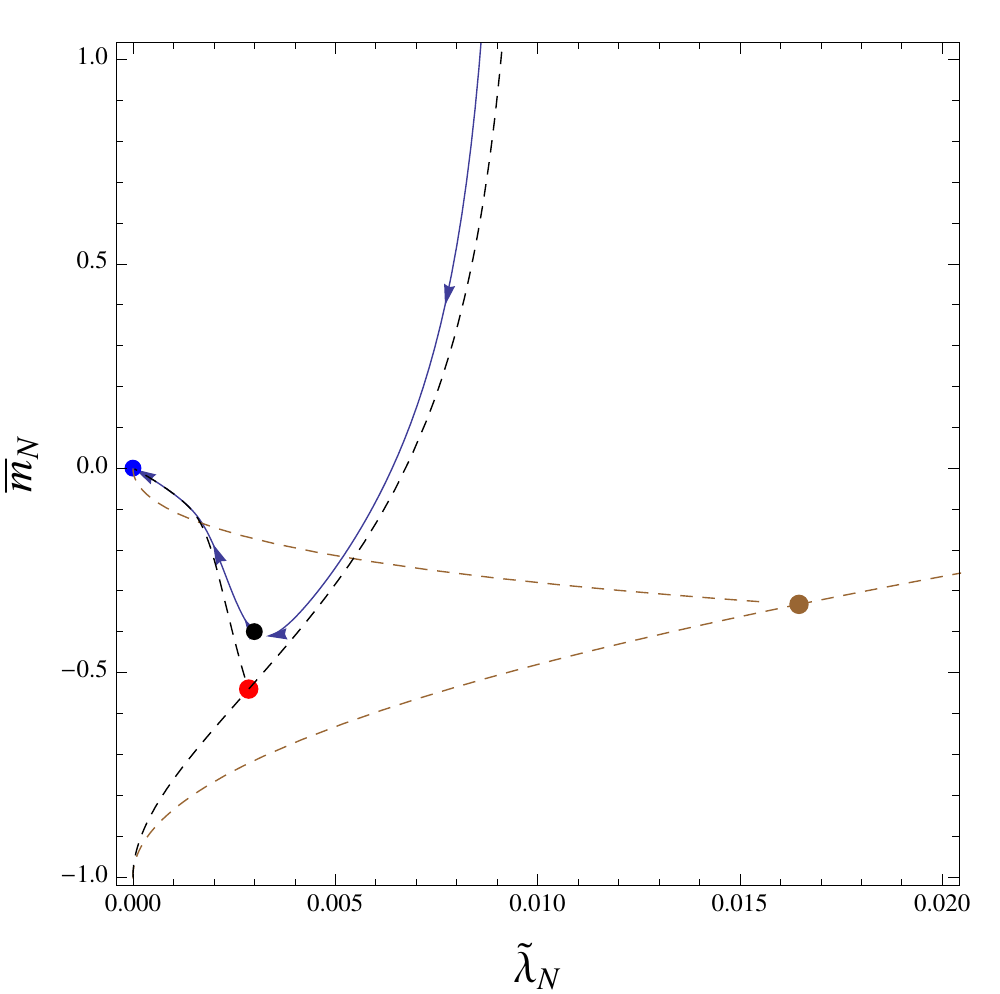} 
\caption{
Flow trajectories (solid blue line) for initial conditions at $\tilde\lambda_{N_0} = 0.003$ and $\bar m_{N_0}=-0.4$ for  (from left to right) $t_0=\ln N_0=0$, $t_0=2$ and $t_0=10$. The black dot locates the initial condition, the blue dot is the GFP, the red dot is the large-$N$ NGFP and the brown one is the small-$N$ NGFP. Black and brown dashed lines denote the eigen-trajectories of the NGFPs.
In the last plot to the right the IR behavior appears to follow the relevant direction at the large-$N$ NGFP, but in truth this is just a long intermediate regime: zooming out the plot one finds that at $\bar m_N\sim 50$ the curve crosses black dashed line, and slowly turns toward the brown one.
}
\label{Fig:flow-fullN}
\end{center}
\end{figure}

In order to study the full system  \eqref{mZ}-\eqref{lambdaZ} numerically,\footnote{We approximate here all the staircase functions with smooth function simply stripping off any floor function from $N$.}
 and be able to recognise the small- and large-$N$ regimes discussed above, it is convenient to introduce a function of $N$ with which to rescale $\lambda_N$ in such a way to interpolate between the two different scalings. This is achieved for example by
\be
\lambda_N = \frac{N^2}{1+N^2} \tilde\lambda_N\, .
\ee
For large $N$, this reduces to an identity, while for small $N$ it reduces to \eqref{lambdabar}.
We show in Figure~\ref{Fig:flow-fullN} some representative curves for general $N$, using such mixed rescaling of $\lambda_N$.
The plot serves to show how the large- and small-$N$ approximations studied before are matched at intermediate $N$, and how initial conditions at the same value of $\bar m$ and $\tilde\lambda$ but different $t$ lead to different behaviors.
%%%%%%%%%%%%
%%%%%%%%%%%%

\begin{figure}[ht]
\begin{center}
\includegraphics[width=5.5cm]{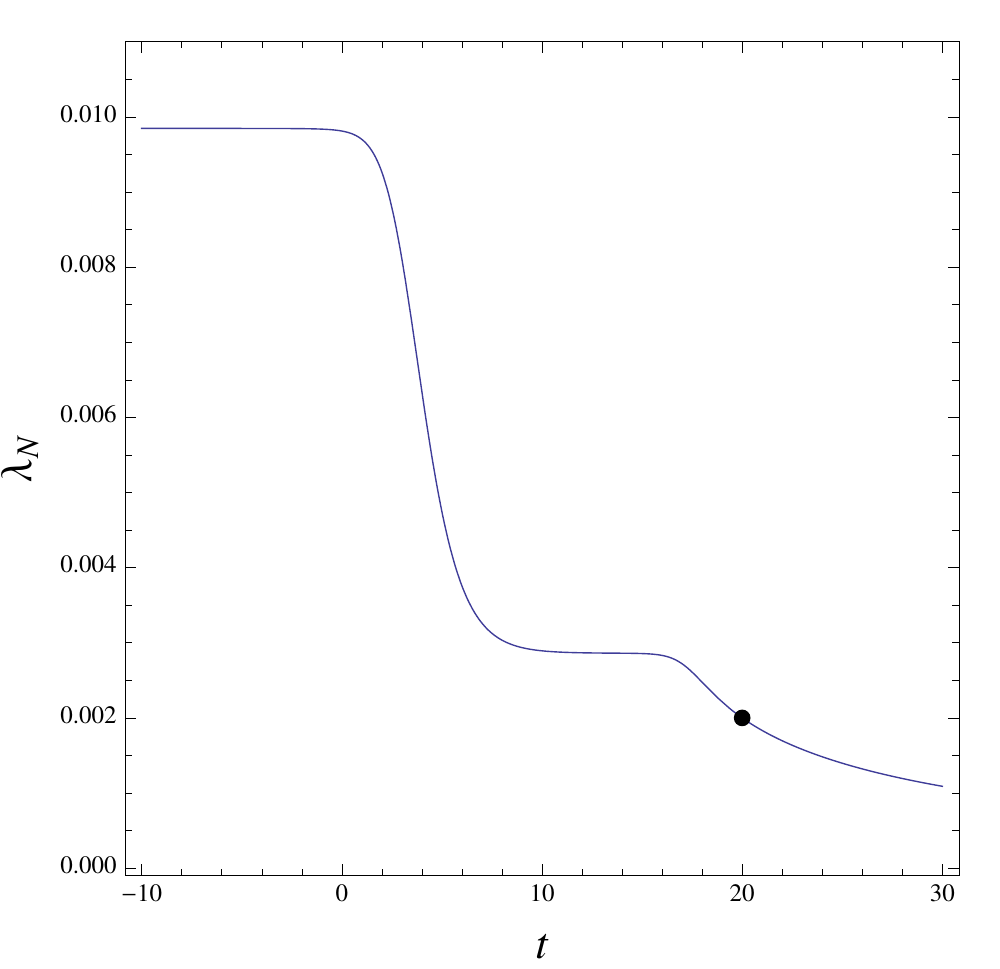} \hspace{.1cm} \includegraphics[width=5.5cm]{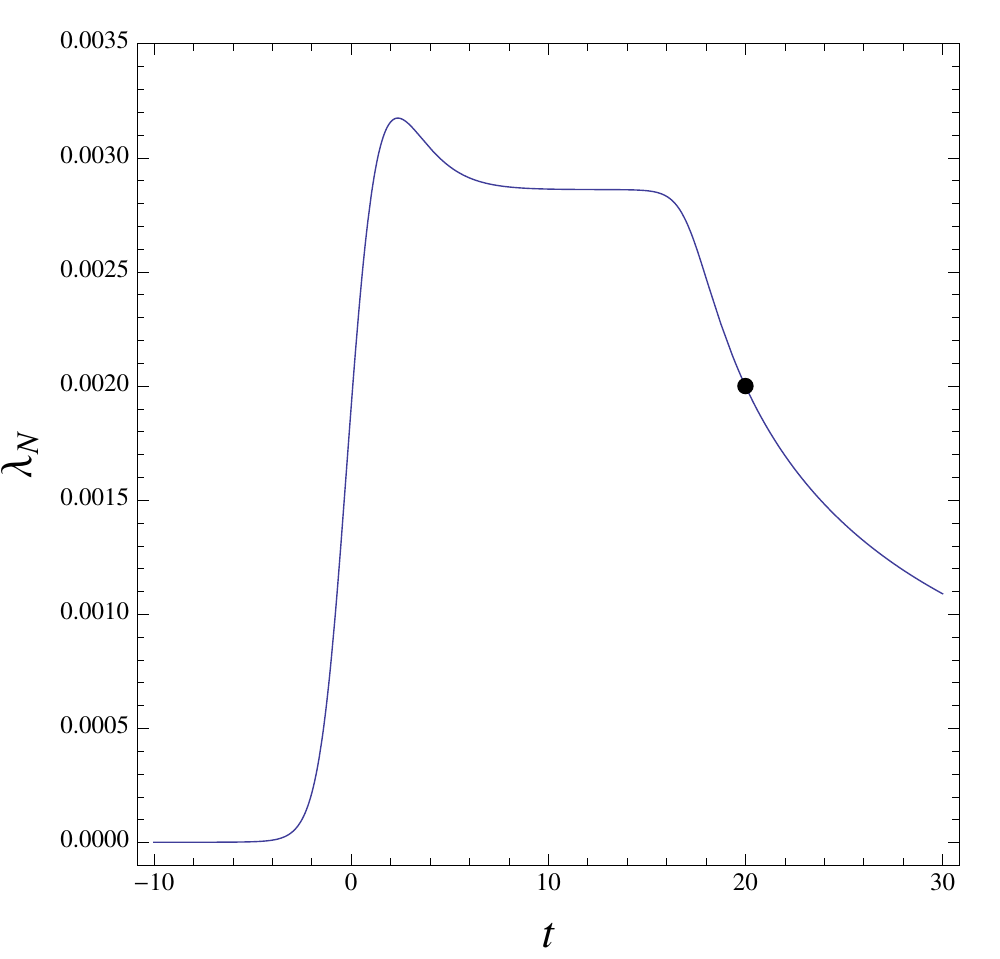}  \hspace{.1cm} \includegraphics[width=5.5cm]{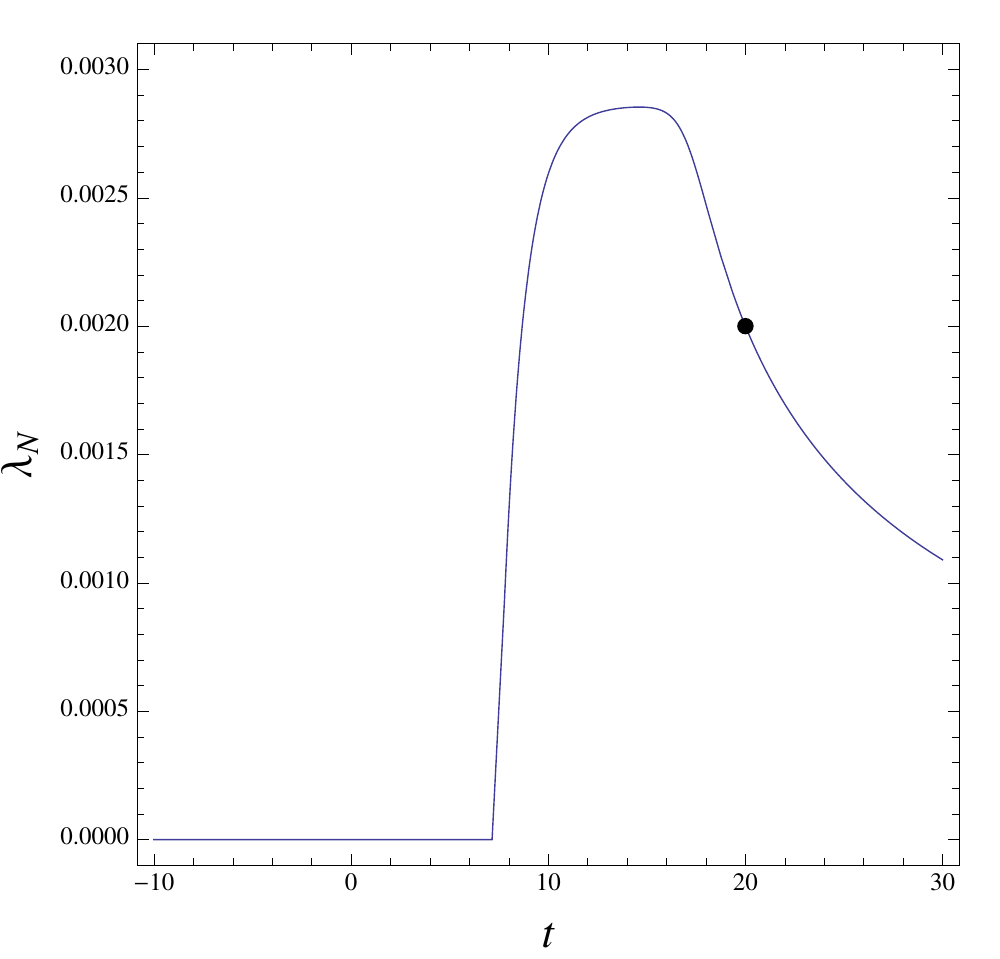} 
\caption{
Flow trajectories for $\lambda_N$ versus $t$, near the critical line. The three curves are for initial conditions (black dot) at $t_0=20$ with $\lambda_{N_0}=0.002$ and (from left to right) $\bar m_{N_0}=-0.22133$, $\bar m_{N_0}=-0.221337043$ and $\bar m_{N_0}=-0.2214$. The curve in the middle is at criticality (the renormalised mass reaches zero with high accuracy).
The flat plateau at $\lambda_N=0$ in the last curve is instead due to the mass hitting the singularity, and it is a manifestation of the bad parametrization in the broken phase, as discussed in the text.
}
\label{Fig:flow-crit}
\end{center}
\end{figure}

It is also useful to study the flow in the original un-rescaled variables. In particular, in this way we can uncover the freezing of the couplings in the deep IR: the value that a coupling takes in such a regime is what we usually call the renormalised coupling, and in terms of that we can usually understand the NGFP as associated to a phase transition. Approaching the phase transition line from above (i.e. from positive initial mass) the renormalised mass goes to zero and then the sign of the mass term changes.  As we already explained, the current parametrization is not ideal for studying the broken phase, but we can certainly study in this way the approach to the critical line. In particular, we can study the behavior of the renormalised coupling for the quartic interaction. For a marginal coupling (such as $\lambda_N$ is at large-$N$) we would expect to find a finite renormalised coupling at criticality if the phase transition is truly there and it is non-trivial. As we already discussed, we expect this to be the case only for large enough $N$, while in the deep IR we expect the renormalised coupling to approach zero as we get close to the critical line.
This is neatly illustrated in Figure~\ref{Fig:flow-crit}, where we show three curves very close to criticality. We find an intermediate regime in which $\lambda_N$ behaves has an IR behavior dictated by the large-$N$ analysis, i.e. it reaches a plateau at $\lambda^*_-$ of \eqref{largeNGFP}. However, going even deeper in the IR, after a short transient, the flow settles in the small-$N$ regime, and in particular we find that as we approach criticality from above $\lambda_N$ reaches a second plateau at zero.
%%%%%%%%%%%%
%%%%%%%%%%%%

%%%%%%%%%%%%%%%
\section{Concluding remarks}
%%%%%%%%%%%%%%%
In this paper, we have developed the general formalism of the functional renormalisation group in the context of (tensorial) group field theories, treating them as standard QFTs on a compact manifolds, but taking into account the non-trivial combinatorics in their interactions, and building up on many results in both GFT \cite{GFTreviews} and tensor models \cite{tensorNew}. We have written down the GFT effective action and the Wetterich equation, incorporating a suitable cut-off function, clarifying in the process the canonical dimensions of GFT fields and couplings and the role of truncations. 
Beside the general equations so obtained, one general lesson is that we obtain a non-autonomous system of beta functions for this type of field theories, due to the presence of an intrinsic scale (the size of the group manifold) in these systems. 

We have then analyzed a concrete example of Tensorial GFT: a 3-dimensional abelian model with linear kinetic term, in its simplest (order 4) truncation. We have derived the equations for the beta functions, forming indeed a non-autonomous system, and given the general properties of the solutions. We analysed in particular its UV (large-cut-off) regime, in which we have found evidence for asymptotic freedom (again for this specific truncation), as well as a non-Gaussian fixed point, and we have identified critical directions of the RG flow. In the IR sector, where we could also solve exactly the RG flow equations, we identified as well the critical points and direction; while we found hints of a phase transition between a symmetric phase and a broken or condensate phase, we have also discussed why these hints are not compelling and instead other indications can be obtained from the RG flow suggesting that no such phase transition exists, without further approximations. This can be attributed mainly to the compact nature of the group manifold defining this GFT model. Indeed, we have also discussed how a further ``flat-regime" approximation of the model gives RG equations that indicate the existence of such a phase transition. We expect other GFT models to show similar features. 

Indeed, we expect our RG analysis of this simple model to provide a useful reference for future work along these lines, on other GFT models of quantum gravity.

We believe that our results, and the FRG in general, will represent an important and powerful tool for the ongoing research on GFT renormalisation \cite{GFTrenorm,josephBeta,Sylvain} and for the study of GFT phase transitions \cite{criticalTensor, GFTphase}, and thus for the outstanding issue of the emergence of continuum spacetime and geometry in quantum gravity \cite{danieleemergence, GFTcondensate, CDTphasetrans, dariojoe}.

\section*{Acknowledgements}
We thank A. Eichhorn, R. Percacci, and especially T. Koslowski for helpful discussions. J.B.G. acknowledges the support of the Alexander von Humboldt Foundation. 

\section*{ Appendix}
\label{app}

\appendix

\renewcommand{\theequation}{\Alph{section}.\arabic{equation}}
\setcounter{equation}{0}

\section{Calculation of the beta functions}
\label{app:wett}

\begin{figure}[h]
\centering
     \begin{minipage}[t]{.8\textwidth}
\begin{center}
\includegraphics[angle=0, width=6cm, height=0.8cm]{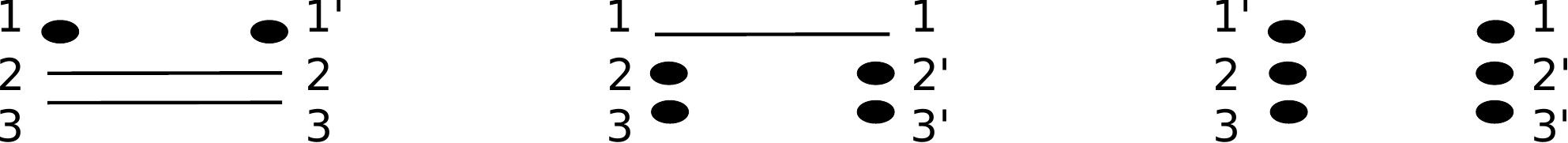}
\caption{ {\small Terms involved in the second derivative
of $\Tr_{4;1}(\varphi^4)$: each link represents an index summation,
each node a un-summed index.}}
 \label{fig:derphi}
\end{center}
\end{minipage}
\end{figure}

In this appendix, we describe the key steps and some interesting
aspects of the calculation of the beta functions \eqref{etafull}, \eqref{mfull}
and \eqref{lambdafull}. 
We start by recalling the expression for $F$, i.e. the contribution to the second derivative of the effective action involving the coupling
$\lambda$: 
\bea
F_N(\{p_i\};\{p_i'\}) 
&=& 
\lambda_N\Big[ \sum_{m_2,m_3}\varphi_{p_1,m_2,m_3} 
\varphi_{p_1',m_2,m_3} \delta_{p_2,p_2'}\delta_{p_3,p_3'}+
\sum_{m_1} \varphi_{m_1,p_2,p_3}\varphi_{m_1,p'_2,p'_3}\delta_{p_1,p_1'}
+   \varphi_{p'_1,p_2,p_3}\varphi_{p_1,p'_2,p'_3}  \cr\cr
&+& \Sym(1\to 2 \to 3) \Big] \,. 
\eea
The first three terms come from the double derivative of $\Tr_{4;1}(\varphi^4)=\sum_{p_i}\varphi_{123} \,\varphi_{1'23} \,\varphi_{1'2'3'} \,\varphi_{12'3'}$. These are depicted in Figure \ref{fig:derphi}, respectively.  

The expansion of the r.h.s. of the  Wetterich equation \eqref{eq:wetter} involves two quantities 
\bea
P_N(\{p_i\};\{p_i'\}) &=& R_N(\{p_i\};\{p_i'\}) +  (Z_N\ft\sum_i |p_i| + m_N)  \bdel_{p_i,p_i'} \,,\cr\cr
F_N(\{p_i\};\{p_i'\}) &=& \Gamma^{(2)}_N(\{p_i\};\{p_i'\})  -   (Z_N\ft\sum_i |p_i| + m_N)  \bdel_{p_i,p_i'} \,.
\eea
and these enter in that expansion as 
\bea\label{app:wet}
 \partial_t \Gamma_n 
 =
 \frac12 \Big[ 
 \overline\Tr\Big(\partial_t R_N \cdot
 P_N^{-1}\Big )+ 
 \sum_{n=1}^{\infty} (-1)^n \overline\Tr\Big(\partial_t R_N \cdot
 P_N^{-1}\cdot[ F_{N} (P_N)^{-1}]^{n}\Big)\Big] .
 \eea

Consider a truncation in the expansion in $n$. 
The first term $\partial_t R_N \cdot
 P_N^{-1}$ collects, as in the ordinary case, vacuum 
configurations and thus can be forgotten. Then, since each term in $F_{N} (P_N)^{-1} $ has exactly two 
 fields, $[F_{N}(P_N)^{-1}  ]^n$ contains $2n$ fields. 
For the flow of $Z_N$ and of the mass $m_N$,
we need to focus on the term of order $n=1$ in \eqref{app:wet}.
These correspond to non-perturbative contributions with two external legs of the form $\varphi^2$. 
Meanwhile, for the running of $\lambda_N$, we must expand at
order $n=2$ and these correspond to terms
with 4 legs, i.e. $\varphi^4$. The other terms do not contribute to the flow, in the chosen truncation.

\bigskip 

\noindent{\bf First order or $\varphi^2$-terms.} 
We introduce the following formal constants:
\bea
c=-\frac12 \,, \qquad
\alpha = -c\frac{\partial_t  Z_N}{(Z_NN+m_N)^2}\,,
\qquad 
\beta =  c\frac{N(\partial_t Z_N +Z_N)}{(Z_N N+m_N)^2} \,.
\eea
The aim is to calculate $c\, \overline{\Tr}\Big(\partial_t R_{N} \cdot (P_N)^{-1}\cdot  [ F_{N}(P_N)^{-1} ]\Big)  = \alpha A+\beta B$. 
Other useful notations must be introduced as this stage:
\bea
&&
\text{for}\, i=1,2,3, \;\; \;\; 
\Tr(\varphi \cdot |p_i|\cdot \varphi ) = \sum_{p_{j}} 
|p_i| (\varphi_{p_j})^2  =  |p_i| \; \inc\;\;, \crcr
&&\text{then}\;\; \;\; 
\Tr (\varphi \cdot K \cdot \varphi)
 = \sum_i |p_i| \; \inc \; = K \;\inc\;, 
\qquad
\Tr(\varphi^2 )\; = \;\includegraphics[angle=0, width=1cm, height=0.5cm]{phi2.pdf}\;\; . 
\eea
We are now in position to evaluate the first order:
 \bea
&&
\alpha A+\beta B = \sum_{p_i} 
\Big[\alpha\Big( \ft\sum_i |p_i|\Big)  + \beta\Big]\Theta(N -  \ft \sum_i |p_i|) F_N(\{p_i\};\{p_i\})\cr\cr
&&= \sum_{p_i} 
\Big[\alpha\Big( \ft\sum_i |p_i|\Big)  + \beta\Big]\Theta(N -  \ft \sum_i |p_i|) \Bigg(\includegraphics[angle=0, width=4cm, height=0.5cm]{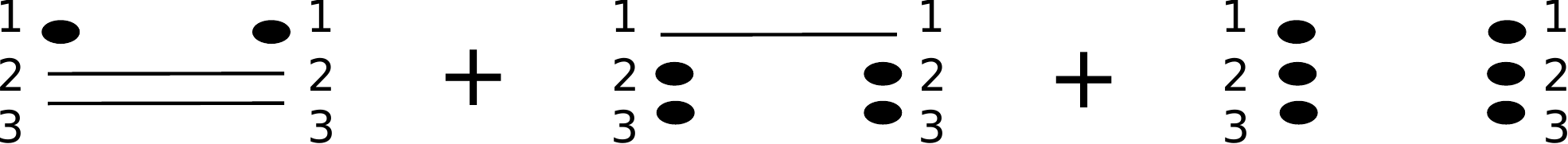} \;+\; \Sym(1\to 2 \to 3)\Bigg) \crcr
&&= 3\;  (\alpha K + \beta)\;  \inc \cr\cr
&& 
+   \Bigg\{ \;\inc\; \Big[
\sum_{p_2,p_3} 
+ \sum_{p_1} 
\Big]\Big[\alpha\Big( \ft \sum_i |p_i|\Big) + \beta\Big]\Theta(N -  \ft \sum_i |p_i|) + \Sym(1\to 2 \to 3) \Bigg\} 
\label{abet}
\eea
The spectral  sums can be performed,  discarding all 
terms involving momenta with power higher than 1, to give
\bea
&&
S_{1;p} = \sum_{p_2,p_3 \in \Z }
\Big[ \alpha\Big( \ft \sum_i |p_i|\Big) + \beta \Big]\Theta(N -  \ft \sum_i |p_i|)  \crcr
&&
 \simeq 4 \sum_{p_2=1}^{3N-|p_1|}\sum_{p_3=1}^{3N-|p_1|-p_2}
\Big[\ft \alpha  \sum_i |p_i| + \beta \Big]
 + 
4 \sum_{p_2=1}^{3N-|p_1|}\sum_{p_3=1}^{3N-|p_1|-p_2}\Big[ 
\ft \alpha( p_2+ |p_1|) + \beta \Big]
+ \ft \alpha |p_1|+ \beta \cr\cr
&& \simeq \frac19\alpha\Big(6N(3N+1)(6N+1) + (1-18N-54N^2)|p_1| \Big) +  \beta \Big(1+6N(1+3 N) -2|p_1|(1+6N)\Big);\crcr
\cr
&&
S_{23;p} =\sum_{p_1 \in \Z} \Big[ \alpha\Big( \ft \sum_i |p_i|\Big) + \beta \Big] \Theta(N -  \ft \sum_i |p_i|) 
\simeq \alpha N(3N+1) + \beta (1+6N  - 2(|p_2|+ |p_3|)) \,.
\eea
Inserting these results in \eqref{abet}, one infers
\bea
&& 
\alpha A+\beta B 
 =3\;  (\alpha K + \beta)\;  \inc \cr\cr
&& 
+ \Bigg\{ \inc \;\Big[ 
\frac19\alpha\Big(6N(3N+1)(6N+1) + (1-18N-54N^2)|p_1| \Big) +  \beta \Big(1+6N(1+3 N) -2|p_1|(1+6N)\Big) \crcr
&&
+ \alpha N(3N+1) + \beta (1+6N  - 2(|p_2|+ |p_3|))  \Big] + \Sym(1\to 2 \to 3)\Bigg\} \crcr
&&
= \inc \;\Bigg\{ 3\;  (\alpha K + \beta)  \cr\cr
&& 
+ 
\frac19\alpha\Big(3\cdot 6N(3N+1)(6N+1) + 3\cdot (1-18N-54N^2)K \Big) +  \beta \Big(3\cdot(1+6N(1+3 N)) -2\cdot3\cdot K(1+6N)\Big) \crcr
&&
+ 3\alpha N(3N+1) + \beta (3(1+6N)  - 2\cdot 6\cdot K)  \Bigg\}
\cr\cr
&&
= \inc\;\Bigg\{\,K \,\Big[ 3\alpha +\alpha \frac19\cdot 3\cdot (1-18N-54N^2)
+ \beta \Big(-2\cdot 3(1+6N)\Big) + \beta (- 2\cdot 6)  \Big] \crcr
&& 
+ \Big[ 3 \beta + \frac19 \cdot 3\cdot 6N(3N+1)(6N+1) 
+\beta  3\cdot (1+6N(1+3 N)) +3\alpha N(3N+1) + \beta 3(1+6N)
  \Big]\Bigg\}\cr\cr
&&
= \inc\;\Big\{ \,K \,\Big[  \alpha(\frac{10}{3}-6N-18N^2)
- 18\beta (2N+1)  \Big] 
+  \Big[\alpha N(3N+1)(12N+5)
+\beta  (9+36N + 54N^2 )
  \Big] \Big\}. \crcr
&& 
\eea
We get the flow equation for $Z_N$ 
and for the mass, after substitution of $\alpha$ and $\beta$ in the previous expression. For $Z_N$ we have 
\bea
\frac{1}{2}\partial_t Z_N = \frac{c \lambda_N}{(Z_N N+m_N)^2} 
\Big[ -\partial_t Z_N (\frac{10}{3} -6N-18N^2)
-N(\partial_t Z_N+Z_N)18(2N +1)
\Big] 
\eea
which easily yields \eqref{etafull}, and the mass term can be calculated as follows
\bea
&&
\frac12 \partial_t  m_N  =   \alpha N(3N+1)(12N+5)
+\beta(9+18N(2+3 N)) \crcr
&&
 = \frac{c \lambda_N}{(Z_NN+m_N)^2}
\Big[
(-\partial_t Z_N) N(3N+1)(12N+5)
+(\partial_t Z_N +Z_N)N(9+18N(2+3 N)) 
\Big] 
\eea
from which $\partial_t m_N$ can be found to be as in \eqref{mfull}.

\bigskip 

\noindent{\bf Second order or $\varphi^4$ terms. }
To compute the flow equation for the coupling constant $\lambda_N$, we have of course to go to higher order in our approximation. We  introduce first
\beq\label{aleph}
c=\frac12 \,, \qquad
 \alpha  = 
- c\frac{\partial_t  Z_N }{(Z_NN+m_N)^2}\,,
\qquad
\beta  = c\frac{N(\partial_t  Z_N + Z_N )}{(Z_N N+m_N)^2}\,.
\eeq
We aim at calculating $c\, \overline{\Tr}\Big(\partial_t R_{N} \cdot (P_N)^{-1}\cdot  [ F_{N}(P_N)^{-1} ]^2\Big)  = \alpha \cA+\beta \cB$. 
This evaluates to 
\bea
&&
\alpha \cA + \beta \cB = 
\sum_{p_i} \Big[ \alpha\Big( \ft \sum_i |p_i|\Big) + \beta \Big] \Theta(N -  \ft \sum_i |p_i|) \sum_{p_i'} F_{N}(\{p_i\};\{p_i'\}) 
 \frac{1}{  R_{N}(p_i')  + Z_N \ft \sum_i |p_i'| + m_N }  F_{N} (\{p_i'\};\{p_i\}) 
\cr\cr
&&  = 
\sum_{p_i} \Big[ \alpha\Big( \ft \sum_i |p_i|\Big) + \beta \Big] \Theta(N -  \ft \sum_i |p_i|) \times \crcr
&& \lambda_N^2
\sum_{p_i'} 
\Bigg[\incPp \,\delta_{p_2,p_2'}\delta_{p_3,p_3'}+\, \incPpp \,\delta_{p_1,p_1'}+\, \incPppp \,+\,  \Sym(1\to 2 \to 3)\Bigg] \frac{1}{  R_{N}(p_i')  + Z_N \ft \sum_i |p_i'| + m_N }  \cr\cr
&& \times 
\Bigg[\incPp \,\delta_{p_2,p_2'}\delta_{p_3,p_3'} +\, \incPpp \,\delta_{p_1,p_1'}+\, \incPppp \,+\,  \Sym(1\to 2 \to 3)\Bigg]\;.
\eea
All these terms define $\varphi^4$ terms. However, not all of them 
will contribute to the flow of $\lambda_N$. 
We will decompose this product according to colors as
$ (A + B + C )P^{-1} (A + B + C ) 
 = (AP^{-1} A + (A\to B \to C)) +
   2(AP^{-1} B + AP^{-1} C +BP^{-1} C)$. 
 We introduce the following notation: 
\bea
P^{-1} (\{p_i'\}) =  \frac{1}{  R_{N}(p_i')  + Z_N \ft \sum_i |p_i'| + m_N } = \wthet_{N-\ft\sum_i p_i'}
\eea 
Then, we focus on the square terms, and get $27 = 3 \times 9$ contributions:
\bea
 &&
 A^2 + (A\to B \to C) =\alpha \cA^1 + \beta \cB^1 = 
 \lambda_N^2 \sum_{p_i} \Big[ \alpha\Big( \ft \sum_i |p_i|\Big) + \beta \Big] \Theta(N -  \ft \sum_i |p_i|) \cr\cr
&& \Bigg\{
 \sum_{p_1'  }\;
\incPp \; \, \incPp \; 
 \wthet_{N-\ft (p_1'+p_2+p_3)} 
\, +  \, \Big(\,\incp \,\Big) \Big(\,\incpp\,\Big)
 \wthet_{-} 
\,+\,\sum_{p_1' }   \,\incPp \; \,\incPHppp \;
 \wthet_{N-\ft (p_1'+p_2+p_3)} 
\cr\cr&&\crcr
&&
+\Big(\,\incpp \,\Big)  \Big(\,\incp \,\Big) 
 \wthet_{-} 
\, + \, \sum_{p_2',p_3'  } \,\incPpp\; \,   \incPpp \; \wthet_{N-\ft(p_1+ p_2'+p_3')}  
\,+\, \sum_{p_2',p_3'  } \,\incPpp\; \,  \incPHIppp\;
\wthet_{N-\ft(p_1+p_2'+p_3')} \cr\cr 
&&
\crcr
&& + \sum_{p_1'} \;\incPHppp\;\, \incPp\;  \wthet_{N-\ft (p_1'+p_2+p_3)} 
\,+\, \sum_{p_2' , p_3' }\; \incPHIppp \;\,\incPpp\; \wthet_{N-\ft(p_1+p_2'+p_3')} 
\, +\,\sum_{p_i'} \Big(\,\incPtwo \,\Big) \Big(\,\incPthree \,\Big) \wthet_{-}   \crcr
&& 
+ 
 \Sym(1\to 2 \to 3) \Bigg\}.
\label{eq:sqa}
\nonumber\eea
There are repeated terms
that we can further reduce in the next step.  
The terms appearing in brackets are disconnected vertices
of the form $(\varphi^2)^2$ and must be neglected. Furthermore, computing 
the spectral sums, since we are only 
considering simple vertices $\Tr_{4;i}(\varphi^4)$,  terms 
like $\Tr_4 (\varphi^4 \cdot p_i^q)$ must not be included in
the flow.
In the following, we will use quite extensively some diagramatic expressions like 
\bea\label{fvert}
\sum_{p_1,p_i'} f(p_1,p_1') \; \vone
\eea
where the dashed lines refer to the fact that we are
summing over indices $p_1$ (or also $p_1'$) called in a
function $f$ and solid lines refer to summed indices 
independent of $f$. Note that the graph gives a particular convolution
of 4 tensors $\varphi^4$.  The goal 
is to provide, in each such summation, the leading term
$f(0,0)$ in a power series expansion of the kernel of the convolution, such that \eqref{fvert}
becomes an ordinary $\Tr_{4;-} (\varphi^4)$ term.

From \eqref{eq:sqa}, we can calculate the next expression:
\bea
&&\alpha \cA^{01} + \beta \cB^{01} =  
 \lambda_N^2 \Bigg\{   \sum_{p_1',p_1  } \vone \sum_{p_2,p_3} \Big[ \alpha\Big( \ft \sum_{i=2,3} |p_i|\Big) + \beta \Big] \Theta_{N -  \ft \sum p}
 \wthet_{N-\ft (p_1'+p_2+p_3)} \Big|_{p_1,p_1'=0}
\crcr
&& 
+\sum_{p_2,p_3, p_2',p_3'  } \;\vexpq 
 \; \sum_{p_1} \Big[ \alpha\Big( \ft  |p_1|\Big) + \beta \Big] \Theta_{N -  \ft \sum_i p_i} 
\wthet_{N-\ft(p_1+ p_2'+p_3')}\Big|_{p_2,p_2',p_3,p_3'=0} \crcr
&& 
+4 \tilde\beta\,  \vex  \; + \;
 \Sym(1\to 2 \to 3) \Bigg\} 
 \eea
 where the last term in $\tilde \beta$ can be calculated in the same
way as explained above and $\tilde\beta = \beta/(Z_NN +m_N)$.
Introducing as well $\tilde\alpha=\alpha/(Z_NN+m_N)$. 
 We get
\bea
\alpha \cA^{01} + \beta \cB^{01} &=&  
 \lambda_N^2 \Bigg\{   \vex
\Bigg[ \sum_{p_2,p_3} \Big[ \tilde\alpha\Big( \ft \sum_{i=2,3} |p_i|\Big) + \tilde\beta \Big] \Theta_{N -  \ft (|p_2|+ |p_3|)}
\crcr
&& 
+  \; \sum_{p_1} \Big[ \tilde\alpha\Big( \ft  |p_1|\Big) + \tilde\beta \Big] \Theta_{N -  \ft |p_1|} 
+4 \tilde\beta \Bigg]  \; + \;
 \Sym(1\to 2 \to 3) \Bigg\} 
\eea
The two internal spectral sums can be calculated as follows.
For some constants $a$ and $b$, one has 
\bea
\tilde{S}_1(N) 
&=&  \sum_{p_2\in \Z}\sum_{p_3\in \Z} 
(a (\ft (|p_2| + |p_3|)) + b) \Theta(N - \ft (|p_2|+|p_3|)) \crcr
&=&
  4\sum_{p_2=1}^{3N} \sum_{p_3=1}^{3N-p_2} 
(\ft a (p_2 + p_3)+ b) + 
4\sum_{p_2=1}^{3N} (\ft a p_2 + b)
 + \beta
\crcr
&=&
 \frac23 a N(3N+1)(6N+1)+ b (1+ 6N+18N^2)
\cr\cr
\tilde{S}_{11}(N)
&=& \sum_{p_1\in \Z} (\ft a |p_1| + b )\Theta(N - \ft|p_1|) \crcr
&=&  2 \sum_{p_1=1}^{3N }  (\ft a p_1 + b)
 + b =  \ft a (3N ) (3N +1) 
+ b (1+  2(3N)  )
 \,\cr\cr
&=& a N (3N +1) 
+ b (1+  6N ) \;. 
\label{specSu}
\eea
Hence we collect the following contributions:
\bea
\alpha \tilde\cA^{01} + \beta \tilde\cB^{01}
&=& \lambda_N^2 \vex\Bigg\{ 
\tilde\alpha \frac{1}{3}N(3N+1)( 12N +5)
+ 6\tilde\beta(1 + 2N + 3N^2)
\Bigg\}+\Sym(1\to 2 \to 3)
\label{eq:b0} \\
&=&  \lambda_N^2 \Bigg[\;\vex \,+\,\secvex\, + \,\trvex\;\Bigg]
 \Bigg\{ 
\tilde\alpha \frac{1}{3}N(3N+1)( 12N +5)
+ 6\tilde\beta(1 + 2N + 3N^2)
\Bigg\}.
\label{id:1}
\eea
Next, we concentrate on the cross-terms and evaluate them
as  (we systematically discard disconnected invariants  in the
following)
\bea
&&2[AB + (AC+ BC)]= \alpha \cA^2 + \beta \cB^2 = 2\lambda_N^2\sum_{p_i} \Big[ \alpha\Big( \ft \sum_i |p_i|\Big) + \beta \Big] \Theta(N -  \ft \sum_i |p_i|)  \times \crcr
&&  
\sum_{p_i'} 
\Bigg[\; \incPp \,\delta_{p_2,p_2'}\delta_{p_3,p_3'}+\, \incPpp \,\delta_{p_1,p_1'}+\, \incPppp  \; \Bigg]\wthet_{N-\ft\sum p'} \cr\cr
&& \times 
\Bigg[\; \intPp\,\delta_{p_1,p_1'}\delta_{p_3,p_3'} +\, \intPpp \,\delta_{p_2,p_2'}+\, \intPppp \;\Bigg]+ \;\; 2(AC+  BC) 
\cr\cr\cr
&& 
=2\lambda_N^2\sum_{p_i} \Big[ \alpha\Big( \ft \sum_i |p_i|\Big) + \beta \Big] \Theta(N -  \ft \sum_i |p_i|)  \times \cr\cr\cr
&&\Bigg\{ 
  \sum_{ p_1'  }   \; \incPp\;\,  \intPHpp\; 
\wthet_{N-\ft(p_1'+p_2+p_3)}
+  \sum_{  p_1' }\; \incPp\;\,
 \incPHppp \;\wthet_{N-\ft(p_1'+p_2+p_3)}
 \cr\cr
&&
+\sum_{ p_2' }\; \intpq \;\; 
\intmp\; 
\wthet_{N-\ft(p_1+p'_2+p_3)}
+ \sum_{ p_3' }\; \intpr \;\, \intppr \;
\wthet_{N-\ft(p_1+p_2+p'_3)}
\cr\cr
&&
+\sum_{ p_2' } \;\intope \;\, \intPp\, \wthet_{N - \ft(p_1 + p_2' + p_3)}
+  \sum_{ p_i' }\; \incPppp \;\, \intPppp\;  
\wthet_{N-\ft(p_1'+ p'_2+p_3')}  
\cr\cr
&&
+  \sum_{ p_2',p_3' }  \incPpp\;\, \intpqr\; 
\wthet_{N-\ft(p_1+ p_2'+p_3')}  
+ \sum_{ p_1',p_3' }\,  \intobe
 \;\,
 \intPpp\; 
\wthet_{N-\ft(p_1'+p_2+p_3')} 
\Bigg\} + 2(AC + BC) \,.
\eea
Summing over the $p_i$ indices, one quickly realizes
that the last two terms in the bracket generate both a 
peculiar connected contribution of the form 
\bea
\incross
\eea
Such terms are in fact well-known since they are 
contraction patterns of a tetrahedron  in the GFT framework \cite{GFTreviews}. 
Thus  tensor invariants can generate this
type of simplexes through the FRGE expansion.  
The converse is also true since tensor invariants
are nothing but  observables obtained from integration
of the color tensor models endowed with precisely such an interaction
\cite{Gurau:2012ix}. For the moment, since we did not
introduce these terms in our truncation, they can not
participate to the flow. It could be however an interesting follow-up
analysis to investigate how their presence in an enlarged truncation 
could lead to modifications of our present results. 

We return to our evaluation and, in a similar way 
as performed above, we put to 0 the momenta
involved in the field convolutions:
 \bea
&&
2[AB + (AC+ BC)]^{0}
=  \cr\cr
&& 
2\lambda_N^2\Bigg\{
\sum_{p_{1},p_{3},p'_1} \vontr 
\sum_{p_2} 
 \Big[ \alpha\Big( \ft  |p_2|\Big) + \beta \Big] \Theta_{N -  \ft \sum_i p_i} \wthet_{N-\ft(p_1'+p_2+p_3)}\Big|_{p_{1},p_{3},p'_1=0}
 \cr\cr
&&
+\sum_{p_2, p_3 ,p_2'} \vosec  
\sum_{p_1}
\Big[ \alpha\Big( \ft |p_1|\Big) + \beta \Big] \Theta_{N -  \ft \sum_i p_i} \wthet_{N-\ft(p_1+p'_2+p_3)}\Big|_{p_{2},p_3,p'_2=0}
\cr\cr
&& 
+  
\tilde\beta \Big(\,\vex +  \secvex +  2\trvex\,\Big) \Bigg\} + 2 (AC + BC)^0\,. 
 \eea
Expanding the sums which are of the form $\tilde{S}_{11}(N)$ \eqref{specSu}, we arrive at 
 \bea
&&
2[AB + (AC+ BC)]^{01}
=  \crcr
&&
2\lambda_N^2\Bigg\{ 
\Bigg[\vex +\secvex\Bigg] (\tilde\alpha N(3N+1) + \tilde\beta (6N+1))
 \cr\cr
&&\;+\;  \tilde\beta \Big(\;\vex +  \secvex + 2\trvex \;\Big)\Bigg\} + 2 (AC + BC)^{01}
 \cr\cr
&& = 
2\lambda_N^2\Bigg\{ \Bigg[\;\vex +  \secvex\;\Bigg](3N+1)(\tilde\alpha N + 2\tilde\beta) + 2 \tilde\beta \trvex\Bigg\}+ 2 (AC + BC)^{01} \crcr
&&
= 2 \cdot 2 \lambda_N^2 \Bigg[\;\vex +  \secvex + \trvex\;\Bigg]
[\tilde\alpha N(3N+1)+ 3\tilde\beta(2N+1)]
\label{id:2}
 \eea 

We are in position to give the flow equation of $\lambda_N$.
Adding \eqref{id:1} and \eqref{id:2}, and after
substituting the value of $\tilde\alpha$ and $\tilde\beta$ 
(with $\alpha$ and $\beta$ given by \eqref{aleph}), one gets:
 \bea
&&
\frac{1}{4}\partial_t  \lambda_N= 
 \frac{\frac12 \lambda_N^2}{(Z_NN+m_N)^3}
 \Bigg\{ 
(-\partial_t Z_N ) \frac{1}{3}N(3N+1)( 12N + 17)
+18(Z_N+\partial_t Z_N)N(1+N)^2 \Bigg\} \crcr
&& 
\partial_t \lambda_N =  \frac{2 \lambda_N^2 N}{(Z_NN+m_N)^3}
 \Bigg\{ 
\partial_t Z_N\frac{1}{3}( 18N^2 +  45N + 37  )
+18Z_N(1+N)^2\Bigg\}
\eea
which gives indeed \eqref{lambdafull}.

%%%%%%%%%%%%%%%
\section{Scaling dimensions}
\label{App:dim}
%%%%%%%%%%%%%%%
In this appendix, we detail the arguments leading to the assignment of canonical scaling dimensions to GFT fields and to the various coupling constants of the model.

Consider the following general action for a rank-$d$ tensor model,
\be
S[\phi] = Z\,\Tr_2(\phi K \phi) + m \,\Tr_2(\phi^2) + g_{{\bf b};n}\,\Tr_{\bf b}(\phi^{n_{\bee}})
\ee
where, as explained in the main text, $K(\{p\};\{\tilde{p}\}) = \bdel_{p_i,p_i'} K(|p_i|)$ is a kernel which will be
summed versus the tensors $\phi_{p}$ and $\phi_{\tilde{p}}$, $m$ is a mass,  ${\bf b}$ is a given pattern of convolution
of the $n_{\bf b}$ tensors, $\Tr_{\bf b}$ sums over tensor
indices which are contracted; $g_{{\bf b},n}$ is a coupling constant
associated with this interaction.  We have removed from 
this expression all symmetry factors, which are irrelevant
for the following discussion. 

We want to assign a canonical scaling dimension to fields and couplings following as much as possible a standard field theory point of view.
Usually this is done (of course in standard units in which $\hbar=c=1$) by ({\it a}) fixing the momentum dimension to $[p]=1$ (from which $[x]=-1$), ({\it b}) imposing $[S]=0$, and ({\it c}) choosing $[Z]=0$ (exploiting the freedom to choose $Z=1$).
Step ({\it a}) is easily implemented: in Tensorial GFTs and tensor models the analogue of momentum modes are the representation labels corresponding to tensor indices, hence it is natural to start with
\be
[N]=[p_i]=1 \, .
\ee
The scaling dimension of fields and couplings will thus correspond to the power of $N$ with which we have to multiply fields and couplings in order to obtain a homogeneous scaling of the action.
The degree of homogeneity is the scaling dimension of the action, which is typically zero (step ({\it b})). We should however be careful on this point, as we should not confuse scaling dimension with engineering dimension. In the absence of external scales, the two notions coincide, but in Tensorial GFTs, as we already pointed out in section \ref{subsect:discNdep}, there is a hidden external scale, the size of the group manifold, here $U(1) \simeq S^1$.
All the would-be-dimensionful quantities (including momenta and running scales) have been made dimensionless by expressing them in units of the radius. As a consequence, the action has of course zero engineering dimension, as it should be. On the contrary, the scaling dimension (i.e. the scaling with $N$) is not fixed a priori by any argument. We can therefore replace step ({\it b}) by the more general assignment
\be
[S]=\alpha\, .
\ee
with $\alpha$ to be chosen on the basis of different considerations at a later stage. Finally, step ({\it c}) applies trivially also in the tensor case.

One more subtlety arises in deciding how to treat summations over field modes. It appears natural to assign scaling dimension $\sigma=1$ to each summation, in analogy with standard momentum integrals in ordinary field theories.
In any case, as we are going to show in a moment, it turns out that the scaling dimension of the couplings is independent of $\sigma$.
We obtain the following system of relations
\be \label{dimeq1}
 [K] +  [\Tr_2]  + 2 [\phi] = \alpha \, ,
\ee
\be \label{dimeq2}
 [m] +  [\Tr_2]  + 2 [\phi] = \alpha \, ,
\ee
\be \label{dimeq3}
 [g_{{\bf b};n}] + [\Tr_{\bf b}] + n_{\bf b}[\phi]    =  \alpha \, .
\ee
Equations \eqref{dimeq1} and \eqref{dimeq2} trivially fix $[m]=[K]$, where $[K]$ is known (from $[p_i]=1$).
Using $[\Tr_2]=d\,\sigma$, then we fix the dimension of the field to
\be
[\phi] =  \frac{\alpha-d\,\sigma-[K]}{2} \, ,
\ee
which, for $\alpha=0$, $\sigma=1$ and $[K]=2$, coincides with the standard dimension of a scalar field in momentum space.
Last, since trace invariants are labelled by $d$-valent graphs  \cite{Bonzom:2012hw}, we deduce that $[\Tr_{\bf b}]=d\,n_{\bf b}\,\sigma/2$, and
\be \label{gn-dim}
[g_{{\bf b};n}]  = \frac{\alpha(2-n_{\bf b})  + n_{\bf b} [K]}{2} \, ,
\ee
which is independent of $\sigma$ as anticipated. Notice that even for $\alpha=0$, $\sigma=1$ and $[K]=2$, the coupling dimension is quite different from the standard one; the origin of the difference is in the fact that here we have $d\,n_{\bf b}/2$ summations over momenta, while for the usual $\phi^n$ interaction we would have $(n-1)d$ integrations.

For $d=3$, and $[K]=\sigma=1$, we have
\be
[\phi] =  \frac{\alpha-4}{2} \, ,
\ee
\be
[m]=1\, ,
\ee
\be
[\lambda]= 2-\alpha\, .
\ee
The scaling dimensions found in section \ref{sect:Flows} by requiring autonomous flow equations correspond to $\alpha=2$ for large $N$
(subsection \ref{subsect:LargeN}) and to $\alpha=0$ for small $N$
(subsection \ref{subsect:smallN}).

\section{Asymptotic freedom of the rank 3 $U(N)$ model}
\label{app:pertur}

As mentionned in previous sections, the $O(N)$ model introduced here is very similar to the $U(N)$ introduced in \cite{BenGeloun:2012pu}. 
Note that in that reference the kinetic term involves a kernel of the
form $(\sum_{s=1}^3 a_s|p_s| + m)$. It is simple to map
at any time $a_s = \frac13 Z$ in order to come back to the 
kinetic term of the model we study here. 

The beta function of the
$U(N)$ rank-3 model has been computed using perturbation theory
at one-loop and one has \cite{BenGeloun:2012pu}: 
\bea\label{ren}
\lambda^{\text{ren}} = \lambda + \lambda^2 S
\eea
where $S$ is a formally divergent sum of the form $S = \sum_{p_i \in \Z}1/[(a(|p_1|+ |p_2|) + m)]^2$. At large $p$'s, truncating this sum at $p\sim \Lambda$ gives a log divergence $\log \Lambda$. 

The perturbative analysis introduced in that reference is a multi-scale analysis \cite{Rivasseau:1991ub}. All
flows are discrete and the momentum shells are discrete
as well.  Summing $S$ between momentum shells between the 
high (UV) slices  $i$ and $i+1$, one arrives at the discrete equation,
counterpart of \eqref{ren} \cite{BenGeloun:2012pu}, 
\bea
\lambda_{i} = \lambda_{i+1} +  \lambda_{i+1}^2 S_{i}\,,  \qquad \qquad 
S_{i} = \frac{4}{a_{i+1}^2} \kappa_M + O(M^{-i})\,. 
\eea 
where $M>1$ and $\kappa_M$ are constant, $a_i$ is the
running of the wave function renormalisation, the rest of the
terms $ O(M^{-i})$ should lead to convergent (suppressed) terms which
should be percluded from the flow. 

In fact, we can even 
complete the calculation of the above reference and give the
exact value of $\kappa_M = 1$.  It turns out that $a_i = a$ is stationary in the UV. 

We decompose the sum defining $S_i$ as  
\be
\begin{split}
 S_i &  = \sum_{M^{i} \leq |p_1| + |p_2| \leq M^{i+1}}\frac{1}{(a(|p_1|+ |p_2|) + m)^2}  \\
& = 4 \sum_{p=M^{i} }^{M^{i+1}} \frac{1}{(a p + m)^2} + 4 \sum_{{\rm max}\{M^{i}-q,1\}}^{M^{i+1}-q} \frac{1}{(a(p+ q) + m)^2} \equiv 4 \mathcal{S}_1 + 4 \mathcal{S}_2 \, .
\end{split}
\ee
The sums $\mathcal{S}_1$ and $\mathcal{S}_2$ can be performed explicitly and written in terms of polygamma functions of order one, and we find that for large $M^i$ they behave as
\be
\mathcal{S}_1 = \frac{1}{a^2 M^i} + O(M^{-i-1}) \, , \;\;\; \mathcal{S}_2 = \frac{1}{a^2} + O(M^{-i}) \, ,
\ee
from which we obtain $\kappa_M = 1$ as promised.

We now rescale all coupling $\lambda_i \to \lambda_i Z^2$. It is easy to 
infer that, at dominant order, 
\bea
\frac{d \lambda(i)}{di} = \beta_\lambda (i)= -  \frac{4}{a^2}\,
Z^2 \lambda_{i+1}^2 \,. 
\eea
Now taking the same convention as in this paper, 
namely $a= \frac13 Z $, we finally get
\beq
\frac{d \lambda(i)}{di}= -36 \, \lambda_{i+1}^2 
\label{ilam}
\eeq
which can be compared to \eqref{betal2} of section \ref{subsect:LargeN}
of the form
\bea
\partial_t \lambda_N = -36 \lambda^2_N\,. 
\label{tlam}
\eea
This is a simple striking relation that it is important to
emphasize. 

However, in order to prove that the $O(N)$ model
is asymptotically free along the lines of the
analysis as performed in that reference,
one must first prove its just-renormalisability 
and then compute  its beta functions. 

We do not perform this detailed analysis, but we stress few interesting facts indicating strongly the
just-renormalisability of the $O(N)$ model of rank $3$. 

The power counting theorem  giving the 
initial divergence degree $\omega_{G}$ of a Feynman amplitude $A_G$, in terms of quasi-local
graphs of the theory, can be achieved following step by step 
Lemma 2, page 1610, in \cite{BenGeloun:2012pu}.  It can be simply 
noticed that the proof of the lemma does not depend on the complex or real nature of the model. In fact, the sole discrepancy that we can 
discern between the real and complex case is given in terms
of the orientability of the graphs. $U(N)$ models have oriented edges
while $O(N)$ models do not. The ``jackets'' which are central in the
proof of Theorem 3, page 1611, might be simply non-orientable
ribbon graphs.
As a consequence, their genus may or may not be an even 
number. But this is a mild modification in the same argument used for the $U(N)$ model, and it is 
not very central for extending the refined power counting illustrated by Theorem  3 to the non-orientable case.
One point, that one should be paying special care to, is 
certainly the graph combinatorics: the $O(N)$ model might be  
not identical of that of the $U(N)$ model (for instance the number
of contractions yielding the one-loop graphs might differ). 
 Perhaps, the similarity of the equations \eqref{ilam} and \eqref{tlam} as illustrated above, might come from the fact that at dominant order  
(large $N$ or large $i$) the graph contributions and their combinatorics 
in these two settings coincide.  This also deserves to be fully analyzed elsewhere.

%%%%%%%%%%%%%%%%%%%%%%%%%%%%%%

\end{document}